\documentclass[aps,twocolumn]{revtex4-1}
\usepackage{times}
\usepackage{graphicx}
\usepackage{amsfonts}
\usepackage{amsmath, amsthm, amssymb}
\usepackage{dsfont}
\usepackage{color}
\usepackage{empheq}
\usepackage{standalone}
\usepackage[most]{tcolorbox}
\usepackage{physics}
\usepackage{bm}
\usepackage[utf8]{inputenc}
\usepackage[margin=0.8in]{geometry}
\usepackage{amssymb,amsmath,amsfonts,amscd}
\usepackage{mathrsfs}
\usepackage{siunitx}
\usepackage{commath}
\usepackage{graphicx} 
\usepackage{braket}
\usepackage{empheq}
\usepackage[most]{tcolorbox}
\usepackage{physics}
\usepackage{placeins}

\usepackage{bibunits}
\defaultbibliography{Refs}
\defaultbibliographystyle{apsrev4-1}

\usepackage{hyperref} 
\hypersetup{
    colorlinks=true,
    linkcolor=blue,    
    citecolor=blue,    
    urlcolor=blue,      
}

\begin{document}

\title{Magnon drag in a metal-insulating antiferromagnet bilayer} 

\author{Eirik Erlandsen}
\affiliation{\mbox{Center for Quantum Spintronics, Department of Physics, Norwegian University of Science and Technology,}\\NO-7491 Trondheim, Norway}

\author{Asle Sudb{\o}}
\email[Corresponding author: ]{asle.sudbo@ntnu.no}
\affiliation{\mbox{Center for Quantum Spintronics, Department of Physics, Norwegian University of Science and Technology,}\\NO-7491 Trondheim, Norway}


\begin{abstract}
We study a bilayer structure consisting of an antiferromagnetic insulator and a normal metal. An electron current is driven in the normal metal with direction parallel to the interface between the materials. Due to interfacial exchange coupling between the localized spins in the antiferromagnet and the itinerant electrons in the normal metal, a magnon current can then be induced in the antiferromagnet. Using an uncompensated antiferromagnetic interface, creating an asymmetry in the interfacial coupling to the two degenerate magnon modes, we find that it is possible to generate a magnon spin-current. The magnon spin-current can be enhanced by increasing the temperature or by spin-splitting the magnon modes.
\end{abstract}

\maketitle

\section{Introduction}
A key element in spin-based electronics is the possibility of using spin-currents to transport information. The spin-currents should be efficiently generated, capable of propagating with low loss of energy, and reliably detected. An interesting avenue for low-loss transportation of spin-signals is provided by magnetic insulators where spin-currents are associated with fluctuations in magnetic order rather than a spin-polarized flow of electrons \cite{Kajiwara2010, Cornelissen2015, Chumak2015, Brataas2020}. Information can thus be transferred without the need of moving charge carriers. Detection of spin-currents propagating through magnetic insulators can e.g.\! be achieved through conversion to electron spin-currents at metal interfaces, which can then be detected through the inverse spin-Hall effect \cite{Hirsch1999, Saitoh2006, Sandweg2011}. Conversely, the generation of spin-currents can be achieved through injection from a neighboring material, such as a material exhibiting the spin-Hall effect \cite{Dyakonov1971, Hirsch1999, Kato2004, Kajiwara2010, Cornelissen2015, Sinova2015, Li2016, Wu2016}. Alternatively, spin-currents in magnetic insulators can also result from e.g.\! a temperature gradient through the spin Seebeck effect \cite{Uchida2008, Uchida2010}.\\
\indent Antiferromagnetic insulators, specifically, have recently gathered interest as alternatives to ferromagnetic insulators as active components in spintronics applications \cite{Jungwirth2016, Baltz2018, Jungfleisch2018, Brataas2020}. An additional complication for spin-transport in antiferromagnetic insulators is, however, that their ability to carry spin-currents can be reduced by competing contributions from the two oppositely polarized magnon modes, often giving rise to a vanishing spin-current for an easy-axis antiferromagnet with two degenerate magnon modes \cite{Ohnuma2013}. Potential solutions to this problem include splitting the magnon modes through e.g.\! the application of an external magnetic field \cite{Rezende2016_2, Wu_2_2016}, or utilizing hard-axis antiferromagnets, naturally featuring non-degenerate magnon modes \cite{Rezende2016}. \color{black} The latter solution relies on the net spin angular momentum of the magnons not vanishing \cite{Ohnuma2013}. \color{black}\\
\indent Going in a different direction, it is also possible to work with degenerate magnon modes, but inducing a magnon spin-current through a coupling to another material where one mode is more affected than the other. Such an asymmetry in the coupling can e.g.\! arise from the other material exhibiting a spin accumulation at the interface \cite{Wang2015, Lin2016, Bender2017, Lebrun2018, Shen2019}, or from the antiferromagnetic interface itself being uncompensated, meaning that only one antiferromagnetic sublattice is exposed at the interface \cite{Bender2017}. In addition to a potential asymmetry in the coupling to the two magnon modes, uncompensated interfaces can also provide an enhancement of electron-magnon interactions through suppressed sublattice interference \cite{Kamra2017, Kamra2019}. This has been exploited in proposals for magnon-mediated superconductivity in heterostructures consisting of antiferromagnets and conductors \cite{Erlandsen2019, Erlandsen2020a, Erlandsen2020b, Thingstad2021}, as well as indirect exciton condensation \cite{Johansen2019}.\\ 
\indent Spin-currents associated with fluctuations in magnetic order can also arise in metallic magnets featuring both ordered localized magnetic moments and itinerant electrons. In this case, the coupling between the localized spins and itinerant electrons can give rise to a rich phenomenology pertaining to transport phenomena \cite{Slonczewski1996, Berger1996, Ralph2008, Tatara2008, Hirohata2020}. For instance, a voltage-induced electron current, naturally giving rise to an electron spin-current in a metallic ferromagnet due to the spin non-degeneracy of the system, can transfer momentum to the magnon population in the system. This gives rise to a magnon spin-current \cite{Cheng2017}. Likewise, a ferromagnetic metal with a temperature gradient will host flow of both electrons and magnons coupled together through drag effects \cite{Blatt1967, Lucassen2011, Miura2012, Yamaguchi2019}. Coupling of flow of electrons and magnons has also been investigated in noncollinear antiferromagnetic metals \cite{Cheng2020}. This type of interplay between electron and magnon currents is not naturally present in magnetic insulators. It can, however, be realized in heterostructures involving magnetic insulators and conducting materials.\\
\indent Recently, it has been proposed that an in-plane charge current carried by spin-triplet Cooper pairs in a superconducting thin-film can induce a magnon spin-current in a neighboring ferromagnetic insulator layer due to interfacial exchange coupling \cite{Johnsen2021}. \color{black} This study, considering the coupling between localized spins and an imbalanced population of left-moving and right-moving particles in an adjacent material, represents a new way of inducing a spin-current in a ferromagnetic insulator. A natural question to ask is then whether it is possible to induce a spin-current in an antiferromagnetic insulator in a similar way. Moreover, as a normal metal subjected to a voltage also can host an imbalance of right-moving and left-moving particles, exchanging the superconductor in Ref.\! \cite{Johnsen2021} with a normal metal could allow for the mechanism to be extended to higher temperatures in a simpler system setup. \color{black}\\
\indent In the present article, we investigate a system consisting of an antiferromagnetic insulator (AFMI) layer located on top of a normal metal (NM) layer, where an in-plane current is driven in the normal metal. Our modelling allows us to tune between a compensated and uncompensated AFMI interface, as well as to introduce spin-splitting of both electrons and magnons. Through interfacial scattering processes, momentum can be transferred from the itinerant electrons of the NM to the magnons in the AFMI, potentially giving rise to magnon currents. Applying semiclassical Boltzmann theory, we here derive a relationship between the macroscopic currents flowing in the system.\\
\indent For the case of spin-degenerate quasiparticles in both subsystems and a compensated antiferromagnetic interface, we find that the charge current in the NM induces a magnon current in the AFMI, but no magnon spin-current as the contributions from the two magnon modes cancel. Applying instead an uncompensated AFMI interface, a magnon spin-current is produced. Interestingly, we find that the magnitude of the induced magnon spin-current is not always maximized for a fully uncompensated interface. A weaker asymmetry in the coupling between the NM and the two AFMI sublattices can actually be more favorable, despite the fact that this weakens the typical strength of the electron-magnon coupling. It is further found that the magnon spin-current increases with temperature and that it can be enhanced by spin-splitting the magnon modes.
\section{Model}\label{sec:model} 
\begin{figure}[t] 
    \begin{center}
        \includegraphics[width=1.0\columnwidth,trim= 0.4cm 0.4cm 0.6cm 0.1cm,clip=true]{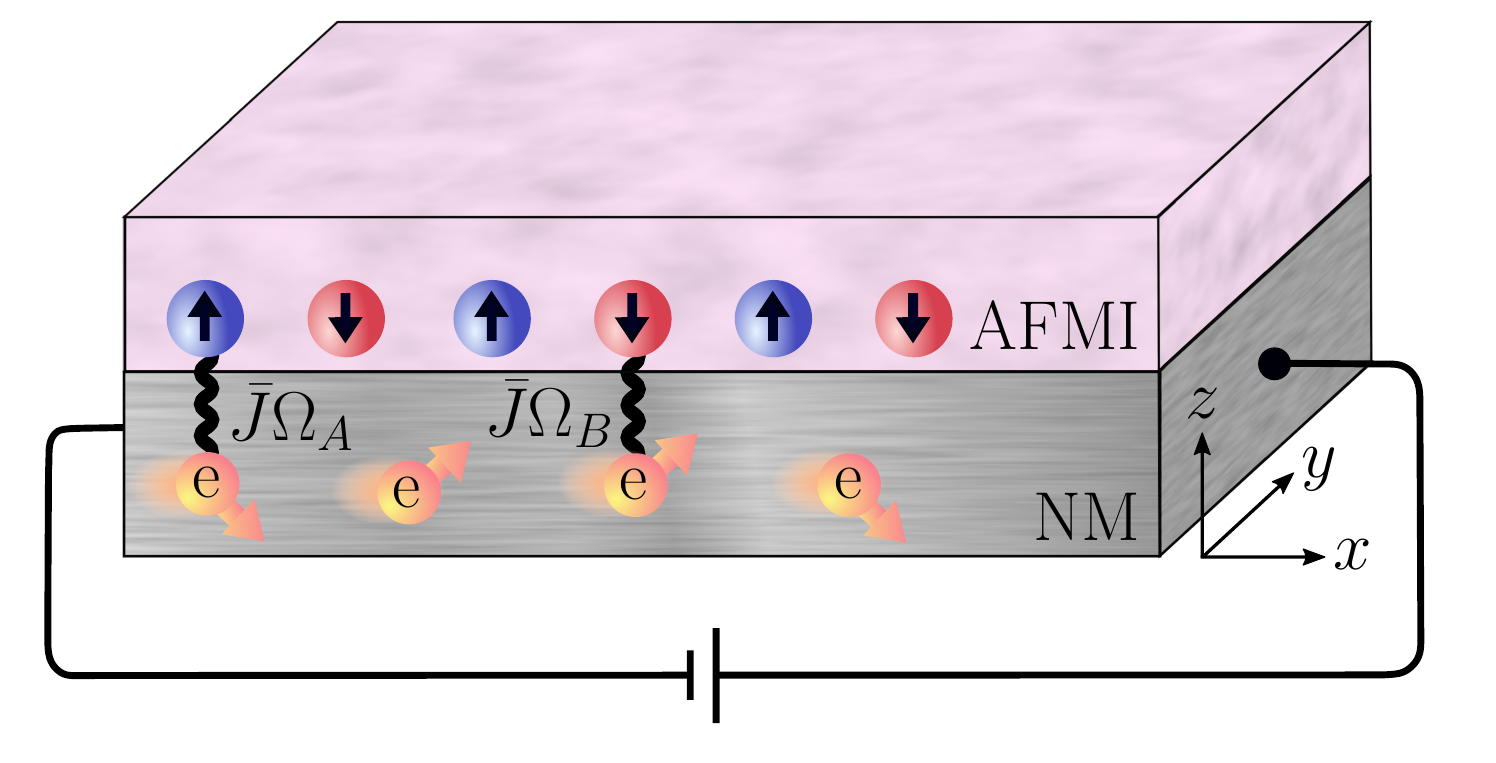}
    \end{center}
    \caption{A bilayer structure consisting of an antiferromagnetic insulator (AFMI) on top of a normal metal (NM). A voltage bias is applied to the normal metal in order to produce an electron current directed along the $x$-axis. The itinerant electrons in the NM can interact with the spins in the AFMI, potentially leading to an induced magnon spin-current. The coupling between the electrons in the NM and the $A$-sublattice of the AFMI is $\bar{J}\Omega_A$, while the coupling to the $B$-sublattice is $\bar{J}\Omega_B$.}
    \label{fig:system} 
\end{figure}

The system setup is illustrated in Fig.\! \ref{fig:system}. An experimental realization of the system will typically feature thin-films of some finite thickness. For simplicity, we consider the layers to be two-dimensional and apply square lattice models. We start out from a tight-binding description of electrons hopping between lattice sites in the NM. For the AFMI, we consider localized spins with easy axis anisotropy $K$, interacting with each other through a nearest-neighbor exchange interaction $J_1$ and a next-nearest neighbor interaction $J_2$. We perform a Holstein-Primakoff transformation in order to describe spin-fluctuations in terms of magnons. Additionally, there is an interfacial exchange coupling $\bar{J} \Omega_{\Upsilon}$ between the localized spins of the $\Upsilon=A,B$ sublattice in the AFMI and the spins of the itinerant electrons in the NM, which gives rise to electron-magnon scattering \cite{Thingstad2021}. Importantly, we can e.g.\! set $\Omega_A = 1$, $\Omega_B = \Omega$ and tune our way from $\Omega = 1$ (compensated interface) to $\Omega = 0$ (uncompensated interface). As discussed in Appendix \ref{sec:App_iso}, we go to the long-wavelength limit to obtain isotropic expressions for the dispersion relations and magnon coherence factors, which will simplify our further calculations. For a sufficiently small and isotropic Fermi surface in the NM, our modelling should be suitable.\\
\indent The Hamiltonian describing the electrons then takes the form

\begin{align}
    H_{\text{NM}} = \sum_{\bm{k} \sigma}\epsilon_{k\sigma}c^{\dagger}_{\bm{k}\sigma}c_{\bm{k}\sigma},\label{eq:H_NM}
\end{align}
where $\epsilon_{k\sigma} = t (ka)^2 - \mu - \sigma h_{e}$. Here, $c^{\dagger}_{\bm{k}\sigma}$ is a creation operators for an electron with momentum $\bm{k}$ and spin $\sigma = \,
\uparrow,\downarrow \,= +,-$. Further, $t$ is the electron hopping amplitude, $a$ is the lattice constant, $\mu$ is the chemical potential, and $h_e$ is a spin-splitting field. The electron spin-splitting can arise from either asymmetric coupling to the two sublattices of the AFMI, an external applied field, or a combination of these two sources, as discussed in Appendix \ref{sec:App_iso}.\\
\indent The Hamiltonian describing the magnons is expressed as  

\begin{align}
    H_{\textrm{AFMI}} = \sum_{\bm{q}}(\omega_{q\alpha}\alpha_{\bm{q}}^{\dagger}\alpha_{\bm{q}} +  \omega_{q\beta}\beta_{\bm{q}}^{\dagger}\beta_{\bm{q}}), \label{eq:H_AFMI_diag}
\end{align}
where $\omega_{q\alpha} = \omega_{q} + h_m$, $\omega_{q\beta} = \omega_{q} - h_m$, and
\begin{align}
    \omega_{q} = \sqrt{\Delta^2_g + \kappa^2 (q a)^2}.
\end{align}
Here, $\alpha^{\dagger}_{\bm{q}}$ is a creation operator for an $\alpha$-magnon (spin down) with momentum $\bm{q}$, and $\beta^{\dagger}_{\bm{q}}$ is a creation operator for a $\beta$-magnon (spin up). The gap in the magnon spectrum is $\Delta_g$, while the dispersiveness of the spectrum is parametrized by $\kappa$. A splitting of the magnon modes $h_m$ could e.g.\! be introduced through an external field. Similarly to the electrons, we will use a short-form notation $\omega_{q,\gamma} = \omega_{q} - \gamma h_m$, where $\gamma = \alpha, \beta = -,+$.\\
\indent Finally, the electron-magnon scattering arising from the coupling between the materials is described by \cite{Thingstad2021}

\begin{align}
    H_{\textrm{int}} = \frac{V}{\sqrt{N}} \sum_{\bm{k}\bm{q}} 
\Big(M_{\bm{q}}\, c_{\bm{k} + \bm{q}, \downarrow}^\dagger c_{\bm{k},\uparrow} 
+ M^{\dagger}_{-\bm{q}}\, c^\dagger_{\bm{k}+\bm{q},\uparrow} c_{\bm{k}, \downarrow}\Big),
\label{eq:H_int}
\end{align}
where $N$ is the number of lattice sites in each layer, $V = -2 \bar{J}\sqrt{S}$, and 

\begin{align}
    M_{\bm{q}} = \big(\Omega_A u_{q} + \Omega_B v_{q}  \big)\alpha_{\bm{q}} + \big(\Omega_A v_{q} + \Omega_B u_{q}  \big)\beta^{\dagger}_{-\bm{q}}.
\end{align}
Here, $S$ is the spin quantum number of the lattice site spins in the AFMI, and the magnon coherence factors $u_{q}$ and $v_{q}$ are defined in Appendix \ref{sec:App_iso}. Importantly, $u_{q}$ and $v_{q}$ have opposite signs and grow large in magnitude when $q \rightarrow 0$, while always satisfying $u^2_{q} - v^2_{q} = 1$. The coupling between electron and long-wavelength magnons can therefore be enhanced by taking $\Omega_A \neq \Omega_B$ \cite{Kamra2019, Erlandsen2019}. In addition to the scattering processes included in \eqref{eq:H_int}, there can also be additional Umklapp scattering processes where the outgoing electron has its momentum shifted by a magnon reciprocal lattice vector \cite{Fjaerbu2019}. Such scattering processes arise because the magnon Brillouin zone is reduced compared to the electron Brillouin zone. For a small electron Fermi surface, Umklapp processes will take electrons far away from the Fermi surface, and such scattering processes can therefore be neglected \cite{Erlandsen2019, Thingstad2021}. Moreover, Umklapp processes will typically not be present for a real uncompensated interface where the NM is lattice matched with one of the two sublattices of the AFMI.\\
\indent In order to describe transport introduced by a voltage bias applied to the NM, we utilize coupled Boltzmann equations for electrons and magnons. We express the linearized Boltzmann equation for the electrons as \cite{Fert1969, Cheng2017}

\begin{align}
       - eE\, v^{e}_{k_x}\frac{\partial f^{0}(\epsilon_{k,\sigma})}{\partial \epsilon_{k,\sigma}} &= - \frac{f_{\sigma}(\bm{k}) - \overline{f_{\sigma}}(k)}{\tau_{\sigma}} - \frac{f_{\sigma}(\bm{k}) - \overline{f_{-\sigma}}(k)}{\tau_{\uparrow\downarrow}}\nonumber\\
       &+ \Bigg[\frac{\partial f_{\sigma}(\bm{k})}{\partial t}\Bigg]_{\textrm{int}}.
\end{align}
Here, $e$ is the elementary charge, $E$ is the electric field applied to the normal metal in the $x$-direction, and $v^{e}_{k_x}$ is the $x$-component of the electron group velocity. Further, $f_{\sigma} (\bm{k})$ is the electron distribution function, $f^{0}(\epsilon_{k,\sigma}) = 1/(e^{\beta \epsilon_{k,\sigma}} + 1)$ is the equilibrium electron distribution function, and $\bar{f}_{\sigma}$ represents a momentum average over the angular coordinate. In the absence of even-in-momentum corrections to the equilibrium distribution, the angularly averaged distribution is equivalent to the equilibrium distribution. Finally, $\tau_{\sigma}$ is the spin-conserving electron relaxation time for electrons with spin $\sigma$, and $\tau_{\uparrow\downarrow}$ is the spin-flip relaxation time for electrons. We have assumed that the electron distribution function is independent of in-plane position. An applied, uniform, electric field gives rise to spatially uniform corrections to the electron distribution functions, giving rise to flow of electrons. Interaction with magnons, represented by the last term \cite{Kasuya1959}, can modify, and potentially spin-polarize, the electron current. These effects are also assumed to be spatially uniform.\\
\indent Furthermore, we express the linearized Boltzmann equation for the magnons as \cite{Rezende2016, Cheng2017}

\begin{align}
        \Bigg[\frac{\partial \,b_{\gamma}(\bm{q})}{\partial t}\Bigg]_{\textrm{int}} = \frac{b_{\gamma}(\bm{q}) - b^{0}_{\gamma}(q)}{\tau_{M,\gamma}(q)}.
\end{align}
Here, the magnon distribution function is denoted by $b_{\gamma}(\bm{q})$, while $b^{0}_{\gamma}(q) = 1/(e^{\beta \omega_{q,\gamma}} - 1)$ is an equilibrium magnon distribution function. Moreover, $\tau_{M,\gamma}(q)$ is a momentum-dependent magnon-relaxation time. While the left-hand-sides of the Boltzmann equations for the electrons contains an external driving term, any net magnon motion will have to result from interaction with the electrons in the metal.\\
\indent The electron and magnon distribution functions appearing in the Boltzmann equations will be expressed as sums of the equilibrium distributions and deviations from the equilibrium distributions on the form \cite{Valet1993, Zhang2012, Cheng2017}

\begin{subequations}
\begin{align}
   f_{\sigma}(\bm{k}) &= f^{0}(\epsilon_{k,\sigma}) - \frac{\partial f^{0}(\epsilon_{k,\sigma})}{\partial \epsilon_{k,\sigma}}\big[\delta \mu^{e}_{\sigma} + g^{e}_{\sigma}(\bm{k})\big], \label{eq:f_exp}\\
   b_{\gamma}(\bm{q}) &= b^{0}(\omega_{q,\gamma}) - \frac{\partial b^{0}(\omega_{q,\gamma})}{\partial \omega_{q,\gamma}}\big[\delta \mu^{m}_{\gamma} + g^{m}_{\gamma}(\bm{q})\big]\label{eq:b_exp}.
\end{align}
\end{subequations}
While $\delta \mu$ represents a uniform shift of the chemical potential, any deviations associated with momentum-dependent corrections to the excitation energies are captured by the functions $\{g^{e}_{\sigma}(\bm{k}), g^{m}_{\gamma}(\bm{q})\}$. The part of these functions which is odd in momentum may generate a net flow of particles and will therefore be of relevance for this study. Moreover, the interaction terms in the electron Boltzmann equations can, using the interaction Hamiltonian together with Fermi's golden rule, be expressed as

\begin{subequations}
\begin{align}
    \Bigg[\frac{\partial f_{\uparrow}(\bm{k})}{\partial t}\Bigg]_{\textrm{int}} &= \frac{2\pi V^2}{\hbar N}\sum_{\bm{q}} [Q_{\alpha}(\bm{k},\bm{q}) - Q^{\textrm{R}}_{\beta}(\bm{k}, \bm{q})],\\
    \Bigg[\frac{\partial f_{\downarrow}(\bm{k})}{\partial t}\Bigg]_{\textrm{int}} &= \frac{2\pi V^2}{\hbar N}\sum_{\bm{q}}[ Q_{\beta}(\bm{k},\bm{q}) - Q^{\textrm{R}}_{\alpha}(\bm{k}, \bm{q})],
\end{align}
\end{subequations}
while the interaction terms in the magnon Boltzmann equations similarly may be expressed as 

\begin{align}
\begin{aligned}
    \Bigg[\frac{\partial b_{\gamma}(\bm{q})}{\partial t}\Bigg]_{\textrm{int}} &= \frac{2\pi V^2}{\hbar N}\sum_{\bm{k}} Q_{\gamma}(\bm{k},\bm{q}).
\end{aligned}
\end{align}
Here, we have defined 

\begin{subequations}
\begin{align}
    Q_{\alpha}(\bm{k},\bm{q}) = &\big(\Omega_A u_{q} + \Omega_B v_{q} \big)^2\delta\big[\epsilon_{\bm{k},\uparrow} + \omega_{\bm{q},\alpha} - \epsilon_{\bm{k} + \bm{q},\downarrow}\big]\nonumber\\
    &\times\Big( \big[b_{\alpha}(\bm{q}) + 1\big]\big[1 - f_{\uparrow}(\bm{k})\big]f_{\downarrow}(\bm{k}+\bm{q})\\
    &- b_{\alpha}(\bm{q})\big[1 - f_{\downarrow}(\bm{k}+\bm{q})\big]f_{\uparrow}(\bm{k})\Big),\nonumber\\
    Q_{\beta}(\bm{k},\bm{q}) = &\big(\Omega_A v_{q} + \Omega_B u_{q} \big)^2 \delta\big[\epsilon_{\bm{k},\downarrow} + \omega_{\bm{q},\beta} - \epsilon_{\bm{k}+\bm{q},\uparrow} \big]\nonumber\\
    &\times \Big(\big[b_{\beta}(\bm{q}) + 1\big]\big[1 - f_{\downarrow}(\bm{k})\big]f_{\uparrow}(\bm{k}+\bm{q})\\
    &- b_{\beta}(\bm{q})\big[1 - f_{\uparrow}(\bm{k}+\bm{q})\big]f_{\downarrow}(\bm{k})\Big),\nonumber
\end{align}
\end{subequations}
\noindent as well as introduced $Q^{\textrm{R}}_{\gamma}(\bm{k},\bm{q})$, which is related to $Q_{\gamma}(\bm{k},\bm{q})$ by sending $\bm{q} \rightarrow -\bm{q}$ followed by sending $\bm{k} \rightarrow \bm{k} + \bm{q}$. We see that, with some necessary relabelling of momentum indices, conservation of spin dictates the structure of the equations. Processes increasing/decreasing the number of $\alpha$-magnons contribute in the same way to the number of spin-$\uparrow$ electrons, and conversely to the number of spin-$\downarrow$ electrons. For the $\beta$-magnons, the situation is the same, except for reversal of the spin-directions. 

\section{Deriving macroscopic equations}
\indent Starting from the coupled Boltzmann equations, we derive a set of macroscopic equations relating the spin-polarized magnon current density $j_{sm}$, magnon current density $j_m$, and spin-polarized electron current density $j_s$ to the electron current density $j_e$. The particle current densities are defined as \cite{Cheng2017, Rezende2016_2, Rezende2018}

\begin{subequations}
\begin{align}
    j_s &= \frac{1}{(2\pi)^2}\int\!\textrm{d}\bm{k}\,v^{e}_{k_x}\big[f_{\uparrow}(\bm{k}) - f_{\downarrow}(\bm{k})\big],\\
    j_e &= \frac{1}{(2\pi)^2}\int\!\textrm{d}\bm{k}\,v^{e}_{k_x}\big[f_{\uparrow}(\bm{k}) + f_{\downarrow}(\bm{k})\big],\\
    j_{sm} &= \frac{1}{(2\pi)^2}\int\!\textrm{d}\bm{q}\,v^{m}_{q_x}\big[b_{\beta}(\bm{q}) - b_{\alpha}(\bm{q})\big],\\
    j_m &= \frac{1}{(2\pi)^2}\int\!\textrm{d}\bm{q}\,v^{m}_{q_x}\big[b_{\beta}(\bm{q}) + b_{\alpha}(\bm{q})\big],
\end{align}
\end{subequations}
where we in the thermodynamic limit have introduced integration over momentum. In the following, $j_s$ and $j_{sm}$ will be referred to as the electron and magnon spin-currents, while $j_m$ and $j_e$ will be referred to as simply the magnon and electron currents. Notably, we consider currents in the $x$-direction in real space, and the spin-space $z$-component of the spin-currents. Further, the electron and magnon velocities appearing in the definitions of the currents are given by

\begin{align}
    v^{e}_{k_x} = \frac{1}{\hbar}\frac{\partial \epsilon_{k}}{\partial k} \,\hat{k} \cdot \hat{x}, \hspace{0.6cm} v^{m}_{q_x} = \frac{1}{\hbar}\frac{\partial \omega_{q}}{\partial q} \,\hat{q} \cdot \hat{x}.
\end{align}
As the velocities are odd under inversion of momentum (odd in the $x$-direction and even in the $y$-direction), only the corresponding odd part of the distributions functions $f_{\sigma}(\bm{k})$ and $b_{\gamma}(\bm{q})$ contribute to the currents. Denoting the odd part of $g^{e}_{\sigma}(\bm{k})$ and $g^{m}_{\gamma}(\bm{q})$ by an index $o$, we can then write

\begin{subequations}
\begin{align}
    \hspace{-0.2cm}j_s = \frac{1}{(2\pi)^2}\!\int\!\textrm{d}\bm{k}\,v^{e}_{k_x}\sum_{\sigma} \sigma\Bigg[\!- \frac{\partial f^{0}(\epsilon_{k,\sigma})}{\partial \epsilon_{k,\sigma}}\Bigg]g^{e}_{\sigma,o}(\bm{k})&,\\
    \hspace{-0.2cm}j_e = \frac{1}{(2\pi)^2}\!\int\!\textrm{d}\bm{k}\,v^{e}_{k_x} \sum_{\sigma}\Bigg[\!- \frac{\partial f^{0}(\epsilon_{k,\sigma})}{\partial \epsilon_{k,\sigma}}\Bigg]g^{e}_{\sigma,o}(\bm{k})&,\\
    \hspace{-0.2cm}j_{sm} = \frac{1}{(2\pi)^2}\!\int\!\textrm{d}\bm{q}\,v^{m}_{q_x}\sum_{\gamma}\gamma \Bigg[\!- \frac{\partial b^{0}(\omega_{q,\gamma})}{\partial \omega_{q,\gamma}}\Bigg]g^{m}_{\gamma,o}(\bm{q})&,\\
    \hspace{-0.2cm}j_m = \frac{1}{(2\pi)^2}\!\int\!\textrm{d}\bm{q}\,v^{m}_{q_x}\sum_{\gamma}\Bigg[\!- \frac{\partial b^{0}(\omega_{q,\gamma})}{\partial \omega_{q,\gamma}}\Bigg]g^{m}_{\gamma,o}(\bm{q})&.
\end{align}
\end{subequations}
\indent The next step is to multiply the electron Boltzmann equations by an electron velocity $v^{e}_{k_x}$ and integrate over momentum $\bm{k}$ \cite{Cheng2017}. Similarly, we multiply the magnon Boltzmann equations by a magnon velocity $v^{m}_{q_x}$ and integrate over momentum $\bm{q}$. Once again, any even-in-momentum corrections to the distribution functions drop out of the equations so that the remaining terms can be expressed in terms of the currents. Adding or subtracting the two equations for the electrons, we end up with  

\begin{subequations}

\begin{align}
    &E\,T_{+} = \frac{1}{2}P_0 \tau^{-1}_e j_s - \frac{1}{2} Y_0\tau^{-1}_e j_e + \big[F_{\uparrow} + F_{\downarrow}\big],\label{eq:e_ele}\\
    &E \,T_{-} = \frac{1}{2}P_0\tau^{-1}_e j_e - \frac{1}{2}Y_0\tau^{-1}_e j_s + \big[F_{\uparrow} - F_{\downarrow}\big]. \label{eq:s_ele}
\end{align}
\end{subequations}
We have here defined $\tau^{-1}_e = \tau^{-1}_{\uparrow} + \tau^{-1}_{\downarrow}$, $P_0 = (\tau_{\uparrow} - \tau_{\downarrow})/(\tau_{\uparrow} + \tau_{\downarrow})$, $Y_0 = 1 + 2\tau_e/\tau_{\uparrow\downarrow}$,

\begin{align}
    T_{\pm} = \frac{- e}{(2\pi)^2}\int \!\textrm{d}\bm{k}\,(v^{e}_{k_x})^2 \Bigg[\frac{\partial f^{0}(\epsilon_{k,\uparrow})}{\partial \epsilon_{k,\uparrow}} \pm \frac{\partial f^{0}(\epsilon_{k,\downarrow})}{\partial \epsilon_{k,\downarrow}}\Bigg],
\end{align}
and

\begin{align}
    F_{\sigma} = \frac{1}{(2\pi)^2}\int \!\textrm{d}\bm{k}\,v^{e}_{k_x}\Bigg[\frac{\partial f_{\sigma}(\bm{k})}{\partial t}\Bigg]_{\textrm{int}}.\label{eq:F}
\end{align}
As $\tau_{\uparrow\downarrow}$ shows up in the equations on the form $1 + 2\tau_e / \tau_{\uparrow\downarrow}$, its effect can be neglected for $\tau_{\uparrow\downarrow} \gg \tau_e$. Similarly, for the magnons, we obtain 

\begin{subequations}
\begin{align}
    &B_{\beta} + B_{\alpha} = \tau^{-1}_{M_0} \,j_{m}\label{eq:m_mag},\\
    &B_{\beta} - B_{\alpha} = \tau^{-1}_{M_0} \,j_{sm} \label{eq:sm_mag},
\end{align}
\end{subequations}
where

\begin{align}
B_{\gamma} = \frac{1}{(2\pi)^2}\! \int \!\textrm{d}\bm{q}\, v^{m}_{q_x}\, \nu(q)\Bigg[\frac{\partial \,b_{\gamma}(\bm{q})}{\partial t}\Bigg]_{\textrm{int}}\label{eq:M}.
\end{align}
We have here neglected the $\gamma$-dependence of the magnon relaxation time \cite{Rezende2016, Rezende2016_2}, and written $\tau_{M}(q) = \tau_{M_0} \nu(q)$, where we take $\nu(q)$ to be on the form $\nu(q) = 1/(1 + \sum_n d_n (qa)^{n})$. Setting some coefficient $d_n$ nonzero, we can then capture the effect of momentum-dependence of the magnon relaxation time. Further, it is worth noting that if both electrons and magnons are spin-degenerate, the left-hand-side of \eqref{eq:sm_mag} can vanish. A natural result would then be a nonzero magnon current, but no magnon spin-current. However, for $\Omega_A \neq \Omega_B$, the asymmetry between $u_{q}$ and $v_{q}$ can give rise to $Q_{\alpha} \neq Q_{\beta}$, producing $B_{\alpha} \neq B_{\beta}$.\\
\indent In order to evaluate the interaction terms $F_{\sigma}$ and $M_{\sigma}$, we insert the expressions for the distribution functions from Eq.\! \eqref{eq:f_exp} and \eqref{eq:b_exp}. We then have 

\begin{subequations}
\begin{align}
    &\hspace{-0.1cm}F_{\uparrow} = \frac{V^2 a^2}{\hbar(2\pi)^3}\!\int \!\textrm{d}\bm{k}\,v^{e}_{k_x}\!\int \!\textrm{d}\bm{q}\, [Q_{\alpha}(\bm{k},\bm{q}) - Q^{\textrm{R}}_{\beta}(\bm{k}, \bm{q})],\label{eq:F_up}\\
    &\hspace{-0.1cm}F_{\downarrow} = \frac{V^2 a^2}{\hbar(2\pi)^3}\!\int \!\textrm{d}\bm{k}\,v^{e}_{k_x}\!\int\!\textrm{d}\bm{q}\, [Q_{\beta}(\bm{k},\bm{q}) - Q^{\textrm{R}}_{\alpha}(\bm{k}, \bm{q})],\label{eq:F_down}\\
    &\hspace{-0.1cm}B_{\gamma} = \frac{V^2 a^2}{\hbar(2\pi)^3}\! \int \!\textrm{d}\bm{q}\, v^{m}_{q_x}\,\nu(q)\!\int\!\textrm{d}\bm{k}\, Q_{\gamma}(\bm{k},\bm{q}),\label{eq:M_gamma}
\end{align}
\end{subequations}
now with
\begin{subequations}
\begin{align}
    &Q_{\alpha}(\bm{k},\bm{q}) = \beta\big(\Omega_A u_{\bm{q}} + \Omega_B v_{\bm{q}}  \big)^2\delta\big[\epsilon_{\bm{k},\uparrow} + \omega_{\bm{q},\alpha} - \epsilon_{\bm{k} + \bm{q},\downarrow}\big]\nonumber\\
    &\times b^{0}(\omega_{\bm{q},\alpha})\big[1 - f^{0}(\epsilon_{\bm{k}+\bm{q},\downarrow})\big]f^{0}(\epsilon_{\bm{k},\uparrow})\Big(\big[\delta \mu^{e}_{\downarrow} - \delta \mu^{e}_{\uparrow} - \delta \mu^{m}_{\alpha} \big]\nonumber\\
    &+ \big[ g^{e}_{\downarrow}(\bm{k}+\bm{q}) - g^{e}_{\uparrow}(\bm{k}) - g^{m}_{\alpha}(\bm{q}) \big] \Big),\label{eq:Q_alpha_last}\\
    &Q_{\beta}(\bm{k},\bm{q}) = \beta\big(\Omega_A v_{\bm{q}} + \Omega_B u_{\bm{q}}  \big)^2\delta\big[\epsilon_{\bm{k},\downarrow} + \omega_{\bm{q},\beta} - \epsilon_{\bm{k} + \bm{q},\uparrow}\big]\nonumber\\
    &\times b^{0}(\omega_{\bm{q},\beta})\big[1 - f^{0}(\epsilon_{\bm{k}+\bm{q},\uparrow})\big]f^{0}(\epsilon_{\bm{k},\downarrow})\Big(\big[\delta \mu^{e}_{\uparrow} - \delta \mu^{e}_{\downarrow} - \delta \mu^{m}_{\beta} \big]\nonumber\\
    &+ \big[g^{e}_{\uparrow}(\bm{k}+\bm{q}) - g^{e}_{\downarrow}(\bm{k}) - g^{m}_{\beta}(\bm{q}) \big] \Big)\label{eq:Q_beta_last}.
\end{align}
\end{subequations}
Here, $\beta = 1/(k_B T)$, where $T$ is the temperature of the system and $k_B$ is the Boltzmann constant. Redefining $\bm{k} \rightarrow - \bm{k}$ and $\bm{q} \rightarrow - \bm{q}$, we see that we, once again, are left with only contributions from terms involving $g^{e}_{\sigma,o}$ and $g^{m}_{\gamma,o}$. The factors $g^{e}_{\sigma,o}$ and $g^{m}_{\gamma,o}$ will be used in order to obtain $F_{\sigma}$ and $B_{\gamma}$ expressed in terms of currents multiplied by some prefactors. Depending on the order of the $g$'s in the square brackets in \eqref{eq:Q_alpha_last} and \eqref{eq:Q_beta_last}, we give each term an index $a = 1,2,3$. Further, we also give terms arising from $F_{\sigma}$ an index $\gamma$ depending on whether they arise from $Q_{\alpha}$ or $Q_{\beta}$. We then have a total of $16$ terms to evaluate: 12 terms $F^{(a)}_{\sigma,\gamma}$ and 6 terms $B^{(a)}_{\gamma}$, giving rise to $F_{\sigma} = \sum_{\gamma,a} F^{(a)}_{\sigma,\gamma}$ and $B_{\gamma} = \sum_{a} B^{(a)}_{\gamma}$. Each term should be expressed in terms of a combination of currents multiplied by some prefactor $\Gamma$. As outlined in Appendix \ref{App:eval}, we achieve this goal by, for each term, first performing one of the two momentum integrals. For each term, we are then left with an integral over $g^{e}_{\sigma,o}$ or $g^{m}_{\gamma,o}$ which can be related to a combination of currents by replacing additional momentum-dependent factors by some characteristic value determined by the rest of the integral. Along the way, we assume that the electron energy scale is much larger than $k_B T$. The energy $k_B T$ is again assumed to be much larger than the typical magnon energies that contribute to the integrals, which we find to typically be a good approximation for our antiferromagnetic magnons living in two dimensions.\\
\indent Inserting the resulting expressions for the interaction terms into Eqs.\! \eqref{eq:m_mag}, \eqref{eq:sm_mag}, \eqref{eq:e_ele}, and \eqref{eq:s_ele}, we obtain 

\begin{align}
\begin{aligned}
    \tau^{-1}_{M_0} \,j_{m} &= \big[\Gamma^{(1)}_{\beta} + \Gamma^{(1)}_{\alpha} - \Gamma^{(2)}_{\beta} - \Gamma^{(2)}_{\alpha}\big]j_e\\
    &+ \big[\Gamma^{(1)}_{\beta} - \Gamma^{(1)}_{\alpha} + \Gamma^{(2)}_{\beta} - \Gamma^{(2)}_{\alpha}\big]j_s\\
    &- \big[\Gamma^{(3)}_{\beta} + \Gamma^{(3)}_{\alpha} \big]j_m - \big[\Gamma^{(3)}_{\beta} - \Gamma^{(3)}_{\alpha}\big]j_{sm},\label{eq:set_1}
\end{aligned}
\end{align}

\begin{align}
\begin{aligned}
    \tau^{-1}_{M_0} \,j_{sm} &= \big[\Gamma^{(1)}_{\beta} - \Gamma^{(1)}_{\alpha} - \Gamma^{(2)}_{\beta} + \Gamma^{(2)}_{\alpha}\big]j_e\\
    &+ \big[\Gamma^{(1)}_{\beta} + \Gamma^{(1)}_{\alpha} + \Gamma^{(2)}_{\beta} + \Gamma^{(2)}_{\alpha}\big]j_s\\
    &- \big[\Gamma^{(3)}_{\beta} - \Gamma^{(3)}_{\alpha} \big]j_m - \big[\Gamma^{(3)}_{\beta} + \Gamma^{(3)}_{\alpha}\big]j_{sm},\label{eq:set_2}
\end{aligned}
\end{align} 

\begin{widetext}

\begin{align}
\begin{aligned}
    E\, T_{+} = & - \Big[\frac{1}{2}Y_0\tau^{-1}_e - \big[\Gamma^{(1)}_{\uparrow,\alpha} + \Gamma^{(1)}_{\downarrow,\alpha} - \Gamma^{(2)}_{\uparrow,\alpha} - \Gamma^{(2)}_{\downarrow,\alpha}  - \Gamma^{(1)}_{\uparrow,\beta} - \Gamma^{(1)}_{\downarrow,\beta} +  \Gamma^{(2)}_{\uparrow,\beta} + \Gamma^{(2)}_{\downarrow,\beta} \big]\Big]j_e\\
    & + \Big[\frac{1}{2}P_0 \tau^{-1}_e - \big[\Gamma^{(1)}_{\uparrow,\alpha} - \Gamma^{(1)}_{\downarrow,\alpha} + \Gamma^{(2)}_{\uparrow,\alpha} - \Gamma^{(2)}_{\downarrow,\alpha} +  \Gamma^{(1)}_{\uparrow,\beta} - \Gamma^{(1)}_{\downarrow,\beta} + \Gamma^{(2)}_{\uparrow,\beta} - \Gamma^{(2)}_{\downarrow,\beta} \big]\Big]j_s \\
    & + \big[\Gamma^{(3)}_{\uparrow,\beta} + \Gamma^{(3)}_{\downarrow,\alpha} - \Gamma^{(3)}_{\uparrow,\alpha} - \Gamma^{(3)}_{\downarrow,\beta} \big]j_{m} + \big[\Gamma^{(3)}_{\uparrow,\beta} - \Gamma^{(3)}_{\downarrow,\alpha} + \Gamma^{(3)}_{\uparrow,\alpha} - \Gamma^{(3)}_{\downarrow,\beta}\big] j_{sm},\label{eq:set_3}
\end{aligned}
\end{align}
and

\begin{align}
\begin{aligned}
    E\, T_{-} = \Big[&\frac{1}{2}P_0\tau^{-1}_e + \big[\Gamma^{(1)}_{\uparrow,\alpha} - \Gamma^{(1)}_{\downarrow,\alpha} - \Gamma^{(2)}_{\uparrow,\alpha} + \Gamma^{(2)}_{\downarrow,\alpha}  - \Gamma^{(1)}_{\uparrow,\beta} + \Gamma^{(1)}_{\downarrow,\beta} +  \Gamma^{(2)}_{\uparrow,\beta} - \Gamma^{(2)}_{\downarrow,\beta} \big]\Big]j_e\\
    - \Big[& \frac{1}{2}Y_0 \tau^{-1}_e + \big[\Gamma^{(1)}_{\uparrow,\alpha} + \Gamma^{(1)}_{\downarrow,\alpha} + \Gamma^{(2)}_{\uparrow,\alpha} + \Gamma^{(2)}_{\downarrow,\alpha} +  \Gamma^{(1)}_{\uparrow,\beta} + \Gamma^{(1)}_{\downarrow,\beta} + \Gamma^{(2)}_{\uparrow,\beta} + \Gamma^{(2)}_{\downarrow,\beta} \big]\Big]j_s \\
     + &\big[\Gamma^{(3)}_{\uparrow,\beta} - \Gamma^{(3)}_{\downarrow,\alpha} - \Gamma^{(3)}_{\uparrow,\alpha} + \Gamma^{(3)}_{\downarrow,\beta} \big]j_{m} + \big[\Gamma^{(3)}_{\uparrow,\beta} + \Gamma^{(3)}_{\downarrow,\alpha} + \Gamma^{(3)}_{\uparrow,\alpha} + \Gamma^{(3)}_{\downarrow,\beta}\big] j_{sm}.\label{eq:set_4}
\end{aligned}
\end{align}
\end{widetext}
The coefficients $\Gamma$ are defined in Appendix \ref{App:coeffs}, and their indices relate them to one of the terms $F^{(a)}_{\sigma,\gamma}$ or $B^{(a)}_{\gamma}$. For a given set of parameters, these coefficients can be determined through numerical integration.\\   
\indent Solving Eq.\! \eqref{eq:set_4} for $E$ and inserting this into Eq.\! \eqref{eq:set_3}, we obtain

\begin{align}
    j_s = A_{e\rightarrow s} \, j_e + A_{m \rightarrow s} \,j_m + A_{sm \rightarrow s} \,j_{sm}, \label{eq:rightarrow_s}
\end{align}
where $A_{e\rightarrow s}$, $A_{m \rightarrow s}$, and $A_{sm \rightarrow s}$ are defined in Appendix \ref{App:expressions}. Inserting the expression in Eq.\! \eqref{eq:rightarrow_s} into Eq.\! \eqref{eq:set_1}, we further obtain 

\begin{align}
    j_m = C_{e \rightarrow m}\, j_e + C_{sm \rightarrow m} \,j_{sm},\label{eq:rightarrow_m}
\end{align}
where the expressions for the coefficients are once again provided in Appendix \ref{App:expressions}.\\
\indent Finally, combining Eqs.\! \eqref{eq:rightarrow_m} and \eqref{eq:rightarrow_s} with Eq.\! \eqref{eq:set_2}, we can obtain an expression for $j_{sm}$ in terms of $j_e$. We can then use this expression to also obtain expressions for $j_m$ and $j_{s}$ in terms of $j_e$.

\section{Results}
The final result of our calculation is 

\begin{align}
    \begin{pmatrix} j_s \\ j_m \\ j_{sm} \end{pmatrix} = \begin{pmatrix} P_s \\ P_m \\ P_{sm} \end{pmatrix} j_e.
\end{align}
Here, the magnon spin-current drag coefficient is

\begin{align}
\begin{aligned}
    P_{sm} &= A_{e\rightarrow sm} + C_{e \rightarrow m}\,A_{m \rightarrow sm} \\
    & + (A_{e \rightarrow s} + C_{e \rightarrow m} \,A_{m\rightarrow s})A_{s\rightarrow sm},
\end{aligned}
\end{align}
where

\begin{subequations}
\begin{align}
    A_{e\rightarrow sm} &=  \big[\Gamma^{(1)}_{\beta} - \Gamma^{(1)}_{\alpha} - \Gamma^{(2)}_{\beta} + \Gamma^{(2)}_{\alpha}\big]/X_{sm},\\
    A_{s\rightarrow sm} &=  \big[\Gamma^{(1)}_{\beta} + \Gamma^{(1)}_{\alpha} + \Gamma^{(2)}_{\beta} + \Gamma^{(2)}_{\alpha}\big]/X_{sm},\\
    A_{m\rightarrow sm} &=  -\big[\Gamma^{(3)}_{\beta} - \Gamma^{(3)}_{\alpha}\big]/X_{sm},
\end{align}
\end{subequations}
and

\begin{align}
\begin{aligned}
    &X_{sm} = \tau^{-1}_{M_0} + \big[\Gamma^{(3)}_{\beta} + \Gamma^{(3)}_{\alpha}\big] +  C_{sm \rightarrow m}\big[\Gamma^{(3)}_{\beta} - \Gamma^{(3)}_{\alpha} \big]\\
    &- (A_{sm \rightarrow s} + C_{sm \rightarrow m}\, A_{m \rightarrow s})\big[\Gamma^{(1)}_{\beta} + \Gamma^{(1)}_{\alpha} + \Gamma^{(2)}_{\beta} + \Gamma^{(2)}_{\alpha}\big].
\end{aligned}
\end{align}
The indices of the coefficients have been chosen to highlight the origin of the different contributions. The role played by the coefficients in Eq.\! \eqref{eq:rightarrow_s} and \eqref{eq:rightarrow_m}, can e.g.\! provide some information about the the origin of the different terms in the numerator of $P_{sm}$. If $A_{e\rightarrow sm}$ is nonzero, there can e.g.\! be a nonzero $j_{sm}$ even if $C_{e\rightarrow m}$ and $A_{e\rightarrow s}$ vanish, meaning that a nonzero numerator of $A_{e\rightarrow sm}$ represents contributions to $j_{sm}$ that can be interpreted as arising directly from $j_e$. The other terms in the numerator of $P_{sm}$ can be interpreted as arising indirectly from $j_e$ via $j_m$ or $j_s$. Similarly, some understanding of the terms in the denominator of $P_{sm}$ can be obtained. For vanishing $C_{sm\rightarrow m}$, $A_{sm \rightarrow s}$, and $\tau^{-1}_{M_0}$, there is still a remaining term $\big[\Gamma^{(3)}_{\beta} + \Gamma^{(3)}_{\alpha}\big]$ in $X_{sm}$, associated with direct conversion of magnon spin-current into electron current.\\ 
\begin{figure}[t] 
    \begin{center}
    \hspace{-0.8cm}
        \includegraphics[width=0.75\columnwidth,trim= 0.5cm 0.6cm 0.8cm 0.3cm,clip=true]{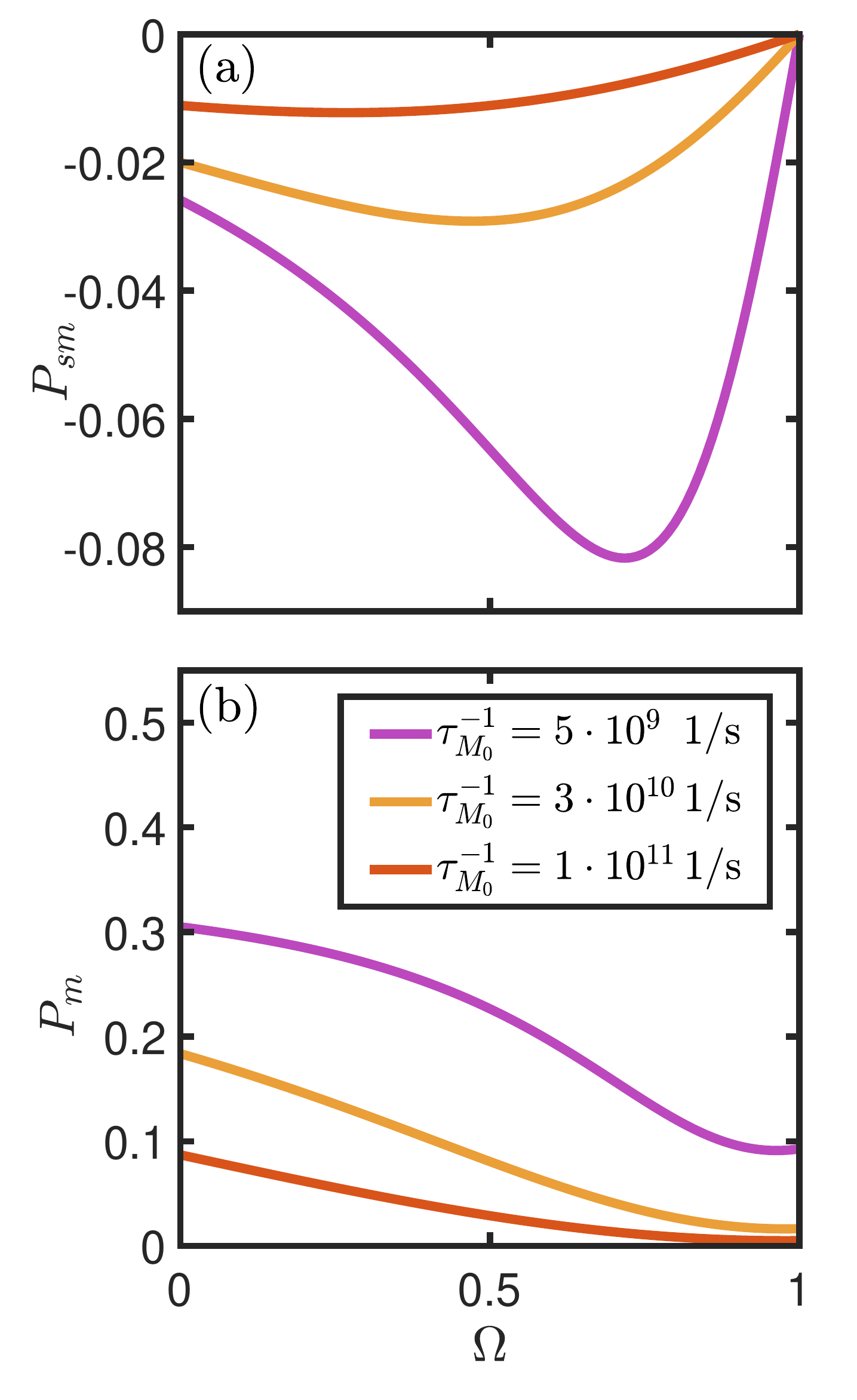}
    \end{center}
        \caption{The ratio of magnon spin-current (a) and magnon current (b) to electron current as a function of the asymmetry in the coupling between the normal metal and the two sublattices of the antiferromagnet. Here, $\Omega = 1$ corresponds to a compensated interface, and $\Omega = 0$ corresponds to an uncompensated interface. Three different curves are displayed for different values of the inverse magnon relaxation time at zero momentum $\tau^{-1}_{M_0}$. The rest of the parameters are set to $t = 1.6\,\textrm{eV}$, $k_F a = 0.6$, $S = 3/2$, $J_1 = 6\,\textrm{meV}$, $J_2 = 0$, $K/J_1 = 1.0\cdot 10^{-3}$, $\bar{J} = 15\,\textrm{meV}$, $T = 300\, \textrm{K}$, $\tau_{\uparrow} = \tau_{\downarrow} = 1.0\cdot 10^{-14}\,\textrm{s}$, $d_3 = 5$, and $h_e = h_m = 0$.}
    \label{fig:vs_Omega} 
\end{figure}
\indent Further, the magnon current drag coefficient can be expressed as

\begin{align}
    P_m = C_{e\rightarrow m} + P_{sm} \,C_{sm\rightarrow m},
\end{align}
and the ratio between the electron spin-current and the normal electron current is

\begin{align}
    P_s = A_{e\rightarrow s} + P_{m}\, A_{m\rightarrow s} + P_{sm}\, A_{sm\rightarrow s} .
\end{align}
The two latter expressions for $P_{m}$ and $P_{s}$ have some room for simplification, but their current form is convenient for understanding the numerical results.\\
\indent Setting $\Omega_{B} = \Omega$ and $\Omega_{A} = 1$, results for $P_{sm}$ and $P_{m}$ as a function of $\Omega$ are presented in Fig.\! \ref{fig:vs_Omega}. We have here neglected any spin-splitting of magnons and electrons and taken $\tau_{\uparrow} = \tau_{\downarrow}$. For $\Omega = 1$, we see that there is a finite induced magnon current, but no magnon spin-current. In order to obtain a magnon spin-current, we need to introduce an asymmetry between the coupling between the electrons and the two magnon modes. This can be achieved by taking $\Omega < 1$, producing a nonzero magnon spin-current. From the figure, we see that $P_m$ simply increases as we reduce $\Omega$. The behavior of $P_{sm}$ is a bit more peculiar. For sufficiently large $\tau^{-1}_{M_0}$, reducing $\Omega$ generally leads to an increase in $|P_{sm}|$ (or at least not a strong reduction), but for smaller $\tau^{-1}_{M_0}$ we see that $|P_{sm}|$ has a clear peak at some $\Omega > 0$. Taking $\Omega = 0$, maximizing the typical strength of the electron-magnon coupling, does, in other words, not necessarily maximize the induced magnon spin-current. We also note that $P_{s}$ is found to be small in all cases.\\
\indent These results can be understood by inspecting the expressions for the drag coefficients. Starting with $P_{sm}$, the behavior is mainly dominated by $A_{e\rightarrow sm}$, where the dominant parts of the denominator of $A_{e\rightarrow sm}$ are the terms related to magnon relaxation and direct conversion of magnon spin-current into electron current. We can then inspect the resulting simplified expression 

\begin{align}
    P_{sm} \sim \frac{\big[\Gamma^{(1)}_{\beta} - \Gamma^{(1)}_{\alpha} - \Gamma^{(2)}_{\beta} + \Gamma^{(2)}_{\alpha}\big]}{\tau^{-1}_{M_0} + \big[\Gamma^{(3)}_{\beta} + \Gamma^{(3)}_{\alpha}\big]}. \label{eq:P_sm_simplified}
\end{align}
Here, asymmetry between $\Gamma^{(1)}_{\gamma}$ and $\Gamma^{(2)}_{\gamma}$ is related to asymmetry between different scattering processes involving a specific magnon mode, arising from an imbalance of electrons moving in opposite directions. Moreover, an asymmetry between $\Gamma^{(a)}_{\alpha}$ and $\Gamma^{(a)}_{\beta}$ has to arise from an asymmetry between scattering processes involving $\alpha$- and $\beta$-magnons. As we see from the simplified expression, asymmetries of both types are necessary in order to obtain a magnon spin-current.\\
\indent Starting from large $\tau^{-1}_{M_0}$, this term dominates the denominator, and the effect of $\Omega$ will enter through the numerator of $P_{sm}$. Here, $\Omega < 1$ will typically act to make $(|u_{q}| - \Omega |v_{q}|)^2$ (for $\alpha$-magnons) and $(|v_{q}| - \Omega |u_{q}|)^2$ (for $\beta$-magnons) larger and more different from each other, increasing the difference between $\alpha$ and $\beta$ contributions to the numerator of $P_{sm}$. We therefore see that $|P_{sm}|$ increases with reduced $\Omega$. As $\alpha$-magnons, for $\Omega < 1$, couple more strongly to electrons due to $|u_q| > |v_q|$, the negative direction of the spin carried by the $\alpha$-magnons makes $P_{sm}$ negative.\\
\indent However, if we reduce $\tau^{-1}_{M_0}$ so that the bracket in the denominator of $P_{sm}$ also starts playing a role, the picture becomes more complicated. In order to have $P_{sm} \neq 0$, we still need $\Omega \neq 1$. Starting from $\Omega = 0$ and increasing $\Omega$, we again have that $\Gamma^{(a)}_{\gamma}$ becomes smaller. Importantly, this reduction mainly stems from suppression of dominant long-wavelength contributions with small $q > 0$. For such contributions $|u_{q}|$ and $|v_{q}|$ are large. Since $|u_{q}|^{2} - |v_{q}|^{2} = 1$, we have that $(|u_{q}| - |v_{q}|) = 1/(|u_{q}| + |v_{q}|))$, meaning that the difference between $|u_q|$ and $|v_q|$ is smaller for long-wavelength magnons. This means that when we increase $\Omega$, we most efficiently suppress the long-wavelength contributions as e.g. \! $(|u_{q}| - \Omega |v_{q}|)^2$ is more efficiently suppressed with increasing $\Omega$ when $|u_{q}|$ and $|v_{q}|$ are very similar. The key point to understanding the behavior of $P_{sm}(\Omega)$ is then that long-wavelength contributions are even more important for the denominator than the numerator of $P_{sm}$. In contrast to the denominator, the numerator, relies on $|u_{q}| \neq |v_{q}|$, not allowing the magnon coherence factors to have their normal boosting effect and making contributions from somewhat larger $q$-values more important. A simple example is $\Omega = 0$. Then, the magnon coherence factors of $\Gamma^{(a)}_{\alpha} - \Gamma^{(a)}_{\beta}$ show up on the form $u^2_{q} - v^2_{q} = 1$, while the coherence factors of $\Gamma^{(3)}_{\alpha} + \Gamma^{(3)}_{\beta}$ show up on the form $u^2_{q} + v^2_{q}$, favoring long-wavelength magnons. By taking $\Omega > 0$, we then strongly suppress the bracket in the denominator through its long-wavelength contributions, while the numerator of $P_{sm}$ is less strongly affected. This makes it possible for $|P_{sm}|$ to increase until the denominator becomes dominated by $\tau^{-1}_{M_0}$, or the suppression of the numerator of $P_{sm}$ due to $\Omega \rightarrow 1$ eventually becomes too strong.\\
\indent A similar increase in $|P_{sm}|$, for sufficiently small $\tau^{-1}_{M_0}$, can also be obtained by reducing the importance of long-wavelength contributions to the $\Gamma$'s in other ways. One option is to increase the easy-axis anisotropy, which both reduces the value of $u_q$ and $v_q$ for small $q$ and increases the excitation energy of long-wavelength magnons. It is, however, worth noting that, within our approximation scheme, one should be careful with suppressing the importance of long-wavelength magnons too much. For this reason, one should also not put too much trust in e.g.\! the results for $P_{m}$ when $\Omega \rightarrow 1$, in contrast to the result $P_{sm}(\Omega = 1) = 0$ which follows from symmetry. Further, taking the magnon relaxation time to decay faster with increasing momentum has an opposite effect, pushing contribution weights towards smaller momenta. If $\tau^{-1}_{M_0}$ is sufficiently small, we can compensate this effect by e.g.\! taking a larger easy-axis anisotropy. However, if $\tau^{-1}_{M_0}$ is too large and/or $\nu(q)$ decays too quickly with momentum, there can be a reduction in the achievable values of $P_{sm}$. Increasing e.g.\! both $d_3$ and $K$ by an order of magnitude, the magnitude of $P_{sm}(\Omega = 0)$ for $\tau^{-1}_{M_0} = 5\cdot 10^{9}\,1/s$ is reduced by around $30\%$.\\ 
\indent Similar to $P_{sm}$, the behavior of $P_m$ in Fig.\! \ref{fig:vs_Omega} can also be understood from a simplified expression

\begin{align}
    P_{m} \sim \frac{\big[\Gamma^{(1)}_{\beta} + \Gamma^{(1)}_{\alpha} - \Gamma^{(2)}_{\beta} - \Gamma^{(2)}_{\alpha}\big]}{\tau^{-1}_{M_0} + \big[\Gamma^{(3)}_{\beta} + \Gamma^{(3)}_{\alpha} \big]}.
\end{align}
We see that the contributions from $\alpha$ and $\beta$ magnons add, in contrast to the case of Eq.\! \eqref{eq:P_sm_simplified} where they were subtracted. In this case, also for small $\tau^{-1}_{M_0}$, there is typically no benefit of increasing $\Omega$ as also the combination of magnon coherence factors in e.g.\! $\Gamma^{(1)}_{\alpha} + \Gamma^{(1)}_{\beta}$ in the numerator still favors long-wavelength magnons. Increasing $\Omega$ then simply leads to a rapid suppression of the numerator.\\
\indent Finally, we also comment on the smallness of $P_s$ in the simple case of no spin-splitting of the electrons. The easiest case to analyze is if we simply take $\tau^{-1}_e$ to be very large. In that case, the denominator of $A_{e\rightarrow s}$, $A_{m\rightarrow s}$ and $A_{sm \rightarrow s}$ contains a term $\tau^{-1}_e$ that is not matched in the numerators. As there, in this case, is assumed to be no intrinsic spin-current source in the normal metal, an electron spin-current will have to arise from interaction with magnons. However, if the electron relaxation time is too short, the effect of interaction with magnons is washed away and the resulting electron spin-current becomes small.\\ 
\begin{figure}[t] 
    \begin{center}
    \hspace{-0.6cm}
        \includegraphics[width=1.0\columnwidth,trim= 0.3cm 1.4cm 0.6cm 0.3cm,clip=true]{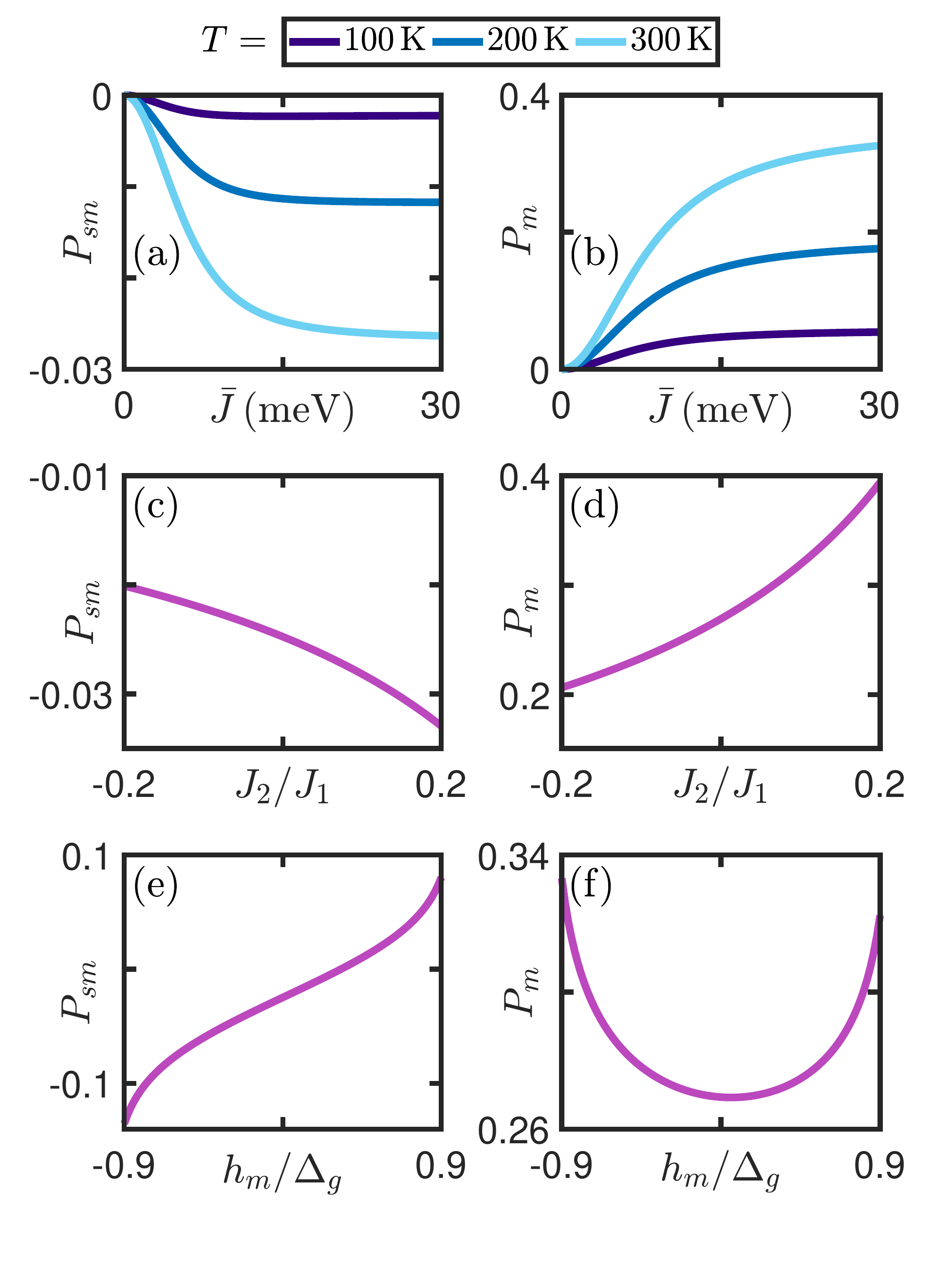}
    \end{center}
    \caption{The ratio of magnon spin-current to electron current, $P_{sm}$, and the ration of magnon current to electron current, $P_{m}$, are here presented. In (a-b) we vary the interfacial coupling between the materials, $\bar{J}$, as well as the temperature $T$. In (c-d) we vary the next-nearest neighbor interaction in the antiferromagnet, $J_2$, where $J_2 > 0$ corresponds to a frustration. In (e-f) we vary the splitting of the magnon modes $h_m$. Unless otherwise specified in the subfigures, we have set $h_m = J_2 = 0$, $\bar{J} = 15\,\textrm{meV}$, and $T=300 \,\textrm{K}$. We have also taken $\Omega = 0$ and $\tau^{-1}_{M_0} =1\cdot 10^{10} \,1/\textrm{s}$, while the rest of the parameters are set to their values in Fig.\! \ref{fig:vs_Omega}.}
    \label{fig:plot2} 
\end{figure}
\indent We next consider, for $\Omega = 0$, how the induced magnon spin-current and magnon current depend on some other important parameters of the system. In Fig.\! \ref{fig:plot2} (a) and (b), we show, for three different temperatures, how $P_{sm}$ and $P_m$ vary with the strength of the interfacial exchange coupling $\bar{J}$. For $\bar{J} = 0$, there is, of course, no induced magnon currents. As all coefficients $\Gamma \sim \bar{J}^2$, increasing $\bar{J}$ makes $\tau^{-1}_{M_0}$ less important until $|P_{sm}|$ and $P_m$ reach their saturation values equivalent to $\tau^{-1}_{M_0} = 0$. The shape of the curves resemble a function $a_1/(1/x^2 + a_2)$, as expected from the simplified expressions for $P_{sm}$ and $P_m$. Further, increasing the temperature leads to more spin-fluctuations, which enter the expressions through the Bose-Einstein distribution factors in $\Gamma^{(1/2)}_{\gamma}$. This increases the magnitude of the numerators of $P_{sm}$ and $P_m$, leading to the results in Fig.\! \ref{fig:plot2} (a) and (b). How much the magnon currents actually can be enlarged by increasing the temperature will necessarily be limited by how large temperature the order in the antiferromagnetic material can survive.
In order to obtain larger drags at a given temperature, one can e.g.\! reduce the energy scale of the antiferromagnet by reducing $J_1$, making it easier to excite magnons. Oppositely, increasing $J_1$ will make it harder to excite magnons and increase the stability of the magnetic ordering. Setting e.g.\! $J_1 = 10\,\textrm{meV}$ in Fig.\! \ref{fig:vs_Omega} leads to $P_{sm}(\Omega = 0) = 0.009$ and $P_{m}(\Omega = 0) = 0.15$ for the purple curves corresponding to $\tau^{-1}_{M_0} = 5\cdot 10^{9}\,1/s$.\\
\indent The behavior of $P_{sm}$ and $P_m$ as a function of next-nearest neighbor interaction in the antiferromagnet is shown in Fig.\! \ref{fig:plot2} (c) and (d). Positive $J_2/J_1$ here corresponds to an antiferromagnetic coupling between next-nearest neighbors, acting as a frustration. Similar to increasing the temperature, frustrating the system leads to more spin-fluctuations producing larger induced magnon currents. Notably, frustration, in contrast to a temperature increase, influences the magnon energies. Frustration therefore also affects the magnon coherence factors, with the particular effect of making them decay more slowly with increasing momentum without affecting their values at zero momentum \cite{Erlandsen2020a, Erlandsen2020b}. The latter effect actually favors the denominator of $P_{sm}$, but the effect of increased number of magnons dominates and makes $P_{sm}$ increase with increasing $J_2$.\\
\indent Finally, in Fig.\! \ref{fig:plot2} (e) and (f), we display how splitting of the magnon modes influences the drag coefficients. Taking $h_m < 0$ lowers the excitation energies of $\alpha$-magnons, increasing the asymmetry favoring contributions to $P_{sm}$ associated with $\alpha$-magnons. This leads to an enhancement of the induced magnon spin-current, which is quite significant because the splitting of the magnon modes allows for long-wavelength magnons to better contribute to $P_{sm}$. Taking $h_m > 0$ works in the opposite direction of the asymmetry between $\alpha$- and $\beta$-magnons introduced by $\Omega = 0$. Moreover, while $P_{m}$ is also influenced by $h_m$, the effect is much weaker as $P_m$ does not rely on an asymmetry between contributions associated with $\alpha$- and $\beta$-magnons. Taking e.g.\! $h_m < 0$, the effect is that the $\alpha$-contributions become larger while the $\beta$-contributions are suppressed. The growth of the $\alpha$ contributions slightly outweigh the decrease in the $\beta$ contributions, leading to a weak enhancement of $P_m$. \color{black} We also note that, if the results are extended to $|h_m|$ even closer to $\Delta_g$, the magnitudes of the drag coefficients continue to grow larger, but they are not found to diverge. \color{black}\\
\indent In order to attempt to describe the induced magnon currents in a real system featuring a sufficiently thick NM layer, neglecting spin-splitting of the electrons arising from $\Omega \neq 1$ (and/or an external field applied to the antiferromagnet) and taking $\tau_{\uparrow} = \tau_{\downarrow}$ might be a reasonable approximation. At least for a thinner NM layer, these effects could, however, play a larger role. We therefore investigate how the drag coefficients depend on $h_e$ and $\tau_{\uparrow} \neq \tau_{\downarrow}$. The relationship between these parameters is not evident, especially when the electron density of states has weak or no energy dependence, and the relationship should be expected to vary strongly with the details of the system. We therefore simply treat $h_e$ and $\tau_{\sigma}$ as independent parameters and display how their separate and combined effects can influence the results.\\ 
\begin{figure}[t] 
    \begin{center}
    \hspace{-0.6cm}
        \includegraphics[width=1.0\columnwidth,trim= 0.7cm 0.0cm 1.0cm 0.3cm,clip=true]{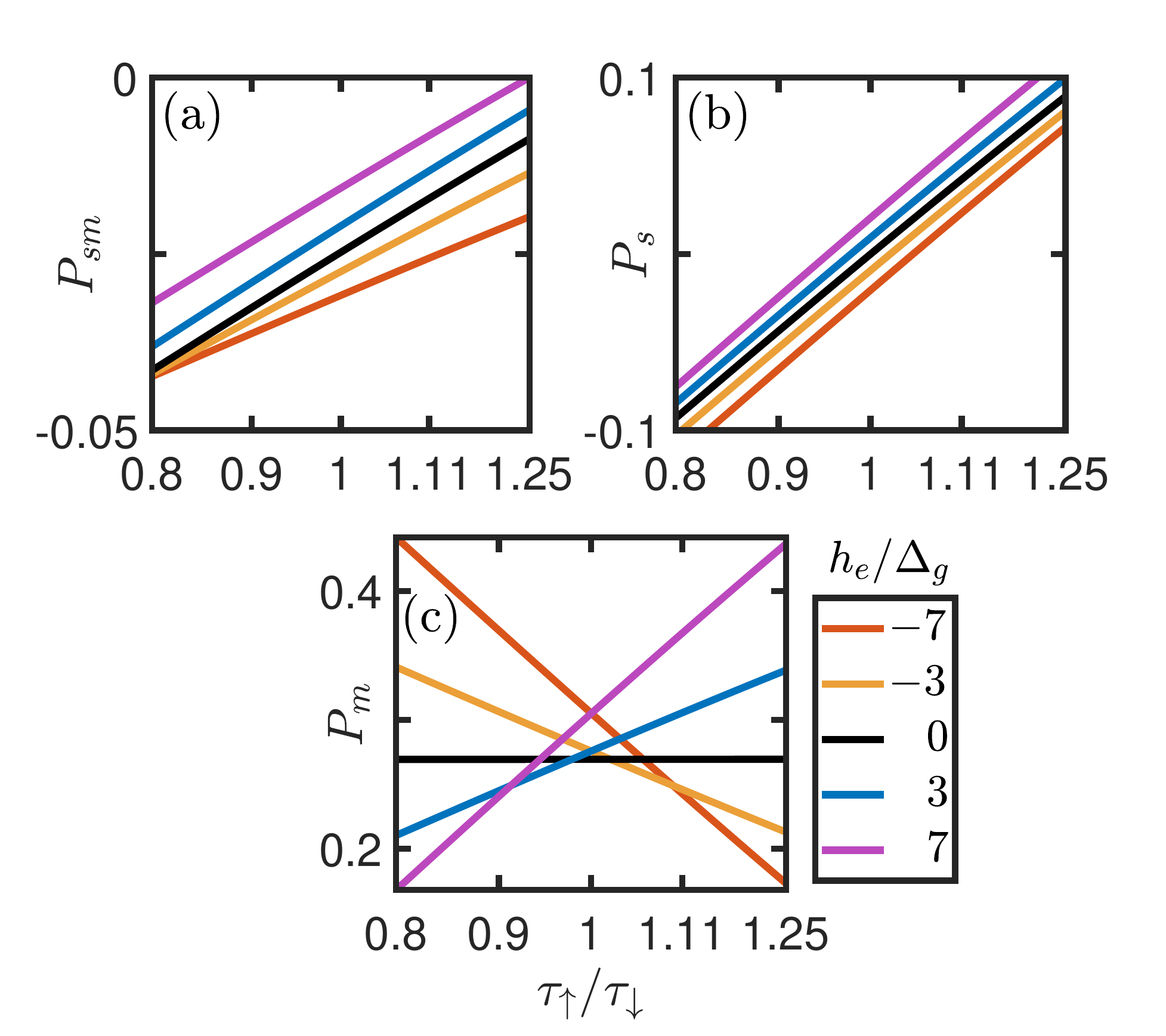}
    \end{center}
    \caption{The ratio of magnon spin-current (a), electron spin-current (b), and magnon current (c) to electron current as a function of $\tau_{\uparrow}/\tau_{\downarrow}$ for different values of $h_e$, where $\tau_{\sigma}$ is the spin-conserving relaxation time for electrons with spin $\sigma$ and $h_e$ represents a spin-splitting of the electron energies. The $x$-axis is here logarithmic in order to highlight symmetries between $\tau_{\uparrow} < \tau_{\downarrow}$ and $\tau_{\uparrow} > \tau_{\downarrow}$. For the parameters, we have set $\Omega = 0$, $\tau_{\downarrow} = 1.0\cdot 10^{-14}\,\textrm{s}$, $\tau^{-1}_{M_0} = 1\cdot 10^{10}\,1/\textrm{s}$, while the rest of the parameters are equal to their values in Fig.\! \ref{fig:vs_Omega}.}
    \label{fig:vs_tausigma} 
\end{figure}
\indent In Fig.\! \ref{fig:vs_tausigma}, we present $P_{sm}$, $P_{s}$, and $P_{m}$ as a function of $\tau_{\uparrow}/\tau_{\downarrow}$ for different values of $h_e$. As we now introduce an asymmetry between spin-$\uparrow$ and spin-$\downarrow$ electrons, we have that e.g.\! $A_{e\rightarrow s}$ starts to grow as $T_{-}/T_{+}$ and $P_0$ can become nonzero. We are then, of course, generating a spin-current, as displayed in Fig.\! \ref{fig:vs_tausigma} (b). For e.g.\! $\tau_{\uparrow} < \tau_{\downarrow}$, the spin-current becomes negative, which is quite natural. Moving on to the magnon spin-current, the last term in the numerator of $P_{sm}$ on the form $(A_{e \rightarrow s} + C_{e \rightarrow m} \,A_{m\rightarrow s})A_{s\rightarrow sm}$ now starts becoming more active. This term can be viewed as converting electron spin-current into magnon spin-current. The general trend is therefore that the changes to $P_{sm}$ in Fig.\! \ref{fig:vs_tausigma} (a) follow the variations in $P_s$. We also see that a spin-splitting field $h_e$ presumably will need to be somewhat larger than $\Delta_g$ in order to have a real effect on the induced magnon spin-current, meaning that the effect on the electrons of applying an external field to split the magnons modes might not be that important for the resulting induced magnon spin-current, even for a thin NM layer. Discussing next the results for $P_{m}$ in Fig.\! \ref{fig:vs_tausigma} (c), we see that $P_m$ can be substantially affected by the combination of splitting of the electron energies and asymmetry in $\tau_{\sigma}$. Taking $\tau_{\uparrow} \neq \tau_{\downarrow}$ alone is enough to create a spin-current through $A_{e\rightarrow s}$, but in order for this spin-current to influence the induced magnon current, we also need a sufficiently large $A_{s\rightarrow m} \sim \big[\Gamma^{(1)}_{\beta} - \Gamma^{(1)}_{\alpha} + \Gamma^{(2)}_{\beta} - \Gamma^{(2)}_{\alpha}\big]$. While the electron spin-current is naturally able to influence the magnon spin-current, we find that e.g.\! the related asymmetry between $\uparrow$ and $\downarrow$ electrons introduced through $h_e$ is needed in order to enhance $A_{s\rightarrow m}$, allowing the electron spin-current to influence the spin-unpolarized magnon current.

\section{Discussion}
As mentioned in the beginning of Sec.\! \ref{sec:model}, an experimental realization of our system would typically involve thin-films of finite thickness. In particular, an experimentally realized uncompensated interface naturally relies on an AFMI with more than one layer. For a given material, the AFMI should be sufficiently thick to, in a real system, stabilize magnetic ordering in the presence of intrinsic quantum and thermal fluctuations, as well as interfacial interaction with the electrons from the metal. While a larger thickness can provide a necessary increase in stability, as well as simply more spins that can fluctuate, the added stability might also make it harder for the electrons of the normal metal to induce magnon currents. The density of the magnon currents should also be expected to decrease as only the surface layer couples directly to the electrons in the normal metal. The present study simply demonstrates that driving an in-plane electron current in an adjacent metal is a potential mechanism for creating a spin-current in an AFMI and highlights that increasing the temperature and splitting the magnon modes are possible ways of increasing the magnon spin-current in an ordered AFMI. In order to obtain reliable estimates for the magnitude of the induced magnon currents in a real system, one would need to take into account the effect of the thickness of the AFMI. Taking properly into account finite thickness of the normal metal layer could potentially also influence the results.\\
\indent Moreover, the model we have used to study the system is, also in other ways, relatively simple, motivated by a goal of exploring some of the key physics that can arise in this system. We have therefore been able to rely on, quite involved, analytical calculations in order to interpret the origin of the results obtainable within the boundaries set by our approximation scheme. A natural extension of this work would be investigations more tailored towards specific material choices. The magnitude of the induced magnon spin-current, intimately related to e.g.\! the competition between effects pushing contribution weights towards smaller or larger momenta, should be expected to depend considerably on the details of the system.\\
\indent For a larger and more anisotropic Fermi surface, Umklapp processes, not considered in this study, could also become of importance. For such scattering processes, $\Omega = 1$ actually maximizes the electron-magnon coupling \cite{Fjaerbu2019, Thingstad2021}. However, an induced magnon spin-current relies on an asymmetry between $\alpha$ and $\beta$ magnons. Hence, in the absence of splitting of the magnon modes, having a compensated interface and relying on Umklapp processes to generate a magnon spin-current does not seem like a viable option.\\ 
\indent As the induced magnon spin-current is found to be enhanced through splitting of the magnon modes, \color{black} antiferromagnets with intrinsically non-degenerate magnon modes present an interesting possibility. Non-degenerate magnon modes that are still able to carry a spin-current may be realized in biaxial antiferromagnets, featuring both a hard-axis and an additional in-plane easy-axis, such as $\textrm{NiO}$ \cite{Rezende2016, Rezende2019}. \color{black} As long as long-wavelength magnons are dominant, it might not be problematic if the splitting of the magnon modes is only significant near the Brillouin zone center. Moreover, as we find that intermediate values of $\Omega$ could be more favorable than $\Omega = 0$, it could be worth considering other options than a fully uncompensated interface. One option could be a compensated interface where the two sublattices are made up of different atoms, potentially introducing an intermediate-strength asymmetry in the coupling between the normal metal and the two sublattices of the AFMI.\\

\section{Summary}

Applying semiclassical Boltzmann theory, we have investigated the possibility of inducing magnon currents in an antiferromagnetic insulator layer through proximity-coupling to a normal metal layer where a charge current is driven parallel to the interface. We find that an asymmetry in the coupling between the electrons and the two sublattices of the antiferromagnet can allow for a magnon spin-current to be generated. The magnitude of the induced magnon spin-current depends intimately on the relative importance of long-wavelength magnons, leading to the somewhat surprising result that a more weakly asymmetric antiferromagnetic interface can be a better choice than a fully uncompensated interface. We also find that the induced magnon currents increase with temperature, and that magnon mode splitting can be beneficial for the magnon spin-current. Future work could include more application-oriented studies, as well as experimentally investigation of our proposed mechanism for generating a magnon spin-current in an antiferromagnetic insulator.    

\section{Acknowledgements}
\indent We thank Even Thingstad for valuable discussions. We acknowledge financial support from the Research Council of Norway Grant No. 262633 ``Center of Excellence on Quantum Spintronics'', and Grant No. 323766 "Equilibrium and out-of-equilibrium quantum phenomena in hybrids
with antiferromagnets and topological insulators".


\appendix

\section{Starting model}\label{sec:App_iso}
Following Ref.\! \cite{Thingstad2021} we start out from a Hamiltonian describing an antiferromagnet

\begin{align}
\begin{aligned}
H_{\textrm{AFMI}} = &J_1 \sum_{ \langle \bm{i},\bm{j} \rangle} \bm{S}_{\bm{i}} \cdot \bm{S}_{\bm{j}} + J_2 \!\sum_{\langle\langle \bm{i},\bm{j}  \rangle\rangle} \bm{S}_{\bm{i}} \cdot \bm{S}_{\bm{j}}\\
&- K \sum_{\bm{i}} S^2_{\bm{i}z} - h_m \sum_{\bm{i}}S_{\bm{i},z}, 
\end{aligned}
\end{align}
where we have added an additional term splitting the magnon modes. Our modelling is not sensitive to whether the spin-space $z$-direction is taken to align with the real-space $z$-direction or not.  Performing a Holstein-Primakoff transformation, this Hamiltonian can be diagonalized and put on the form of Eq.\! \eqref{eq:H_AFMI_diag} with \cite{Thingstad2021}

\begin{equation}
\omega_{\bm{q}} = \sqrt{C_{\bm{q}}^2 - D_{\bm{q}}^2},
\end{equation}
where 

\begin{align}
C_{\bm{q}} &= 2 z_1 J_1 S - 2 z_2 J_2 S ( 1 - \tilde{\gamma}_{\bm{q}} ) + 2KS, 
\\
D_{\bm{q}} &= 2 z_1 J_1 S \gamma_{\bm{q}}.
\end{align}
Here, 

\begin{align}
&\gamma_{\bm{k}} = \frac{2}{z_1}\big[\cos(k_{x} a) + \cos(k_{y} a)\big],\\
&\tilde{\gamma}_{\bm{k}} = \frac{2}{z_2}\big[\cos(k_x a + k_y a) + \cos(k_x a - k_y a)\big], 
\end{align}
where $z_1$ is the number of nearest neighbors, and $z_2$ is the number of next-nearest neighbors.\\
\indent For the coupling between the electrons and magnons, we start out from 

\begin{align}
    H_{\textrm{int}} = -2\Bar{J}\sum_{\Upsilon}\sum_{\bm{i}\in \Upsilon}\Omega_{\Upsilon}\,c_{\bm{i}}^{\dagger}\bm{\sigma}c_{\bm{i}}\cdot \bm{S_i},
\end{align}
where the sum over $\Upsilon \in \{A,B\}$ is a sum over the two sublattices of the AFMI, and $\bm{\sigma}$ is a vector of Pauli matrices. We then, again, perform a Holstein-Primakoff transformation, as well as Fourier transformations. Neglecting Umklapp scattering processes and moving electron spin-splitting terms to the NM Hamiltonian, we arrive at the expression in Eq.\! \eqref{eq:H_int} \cite{Thingstad2021}. Here, the magnon coherence factors, relating the AFMI eigenexcitations to the original sublattice spin-flip magnons introduced in the Holstein-Primakoff transformation, take the form

\begin{align}
    u_{\bm{q}} = \frac{1}{\sqrt{2}}\sqrt{\frac{C_{\bm{q}}}{\omega_{\bm{q}}} + 1},\\
    v_{\bm{q}} = \frac{-1}{\sqrt{2}}\sqrt{\frac{C_{\bm{q}}}{\omega_{\bm{q}}} - 1}.
\end{align}
\indent Finally, for the electrons, we start out from 

\begin{align}
\begin{aligned}
    H_{\textrm{NM}} =  &-t\!\sum_{\langle \bm{i}, \bm{j} \rangle\sigma} c^{\dagger}_{\bm{i}\sigma}c_{\bm{j}\sigma} - \mu \sum_{\bm{i}\sigma}c^{\dagger}_{\bm{i}\sigma}c_{\bm{i}\sigma}\\
    &- h'_e \sum_{\bm{i}\sigma}\sigma c^{\dagger}_{\bm{i}\sigma}c_{\bm{i}\sigma},
\end{aligned}
\end{align}
where $h'_e$ is a spin-splitting arising from an externally applied field. Diagonalizing the Hamiltonian and including the potential additional spin-splitting of the electrons arising from the antiferromagnetic interface, we end up with the Hamiltonian in Eq.\! \eqref{eq:H_NM}. Here,

\begin{align}
    \epsilon_{\bm{k}\sigma} = -t z_1\gamma_{\bm{k}} - \mu - \sigma h_e.
\end{align}
The additional spin-splitting of the electrons arising from the interaction term has a strength $\bar{J}S(\Omega_B - \Omega_A)$. However, if we consider a NM of finite thickness, delivering a similar surface current affecting the antiferromagnet, the effective spin-splitting of the electrons due to proximity to the AFMI will be reduced. We therefore treat the spin-splitting of the electrons as an adjustable parameter.\\ 
\indent Finally, the last step is to consider the long-wavelength limit in order to obtain isotropic expressions for $\omega_{q}$, $\epsilon_{k\sigma}$, $u_{q}$, and $v_{q}$. For $\omega_{q} = \sqrt{\Delta^2_g + \kappa^2 (qa)^2}$, we have defined $\Delta_g = 2 S \sqrt{K(K + 8J_1)}$, and $\kappa = 4S\sqrt{2 J^2_1 - J_2(K + 4 J_1)}$. Further, the magnon coherence factors now take the form 

\begin{align}
    u_{q} = \frac{1}{\sqrt{2}}\sqrt{\frac{8J_1 S - 4J_2 S(qa)^2 + 2KS}{\omega_q} + 1},
\end{align}

\begin{align}
    v_{q} = -\frac{1}{\sqrt{2}}\sqrt{\frac{8J_1 S - 4J_2 S(qa)^2 + 2KS}{\omega_q} - 1}.
\end{align}
As we are working in the long-wavelength limit, under the assumption that large-momentum processes are negligible, it will not be of importance that the magnons live in a reduced Brillouin zone compared to the electrons. Momentum integrals for both magnons and electrons will be performed over circular Brillouin zones of radius $\pi/a$, where we make sure that contributions from large momenta have little or no influence on the results. In particular, by taking a sufficiently small Fermi surface, we make sure that all integrands containing magnon coherence factors vanish before the isotropic expression for $v_{q}$ turns imaginary.

\section{Evaluating the interaction terms} \label{App:eval}
As discussed above, in order to evaluate the interaction terms, we start out from Eq.\! \eqref{eq:F_up}, \eqref{eq:F_down}, \eqref{eq:M_gamma} and divide these expressions up into terms involving a single factor $g^{e}_{\sigma,o}$ or $g^{m}_{\gamma,o}$. For terms involving  $g^{e}_{\sigma,o}(\bm{k} + \bm{q})$ or $g^{m}_{\sigma,o}(\bm{k} + \bm{q})$, we send $\bm{k} \rightarrow \bm{k}-\bm{q}$. We then proceed to do the angular part of the integral over the momentum that the involved factor $g$ does not depend on. If $g$ depends on $\bm{q}$, we can next do the radial integral over $\bm{k}$ and use the remaining integral over $\bm{q}$ to form a combination of magnon currents after having replaced some additional $\bm{q}$-dependent factors by characteristic values. If $g$ depends on $\bm{k}$, we use the assumption that the important magnon energies are significantly smaller $k_B T$ in order to decouple the two integrals. Replacing some additional $\bm{k}$-dependent factors by characteristic values, we are then left with a combination of electron currents with a prefactor that depends on a radial integral over $q$.\\
\indent In this Appendix we outline the evaluation of two specific terms. The rest of the terms can be evaluated in a similar manner. We start with $B^{(2)}_{\alpha}$, which can be written out as

\begin{align}
\begin{aligned}
    &B^{(2)}_{\alpha} = -\beta\frac{V^2 a^2}{\hbar(2\pi)^3}\!\int\!\textrm{d}\bm{k}\,f^{0}(\epsilon_{\bm{k},\uparrow})g^{e}_{\uparrow,o}(\bm{k})\\
    &\times\!\int \!\textrm{d}\bm{q}\, v^{m}_{q_x} \nu(q)\big(\Omega_A u_{\bm{q}} + \Omega_B v_{\bm{q}}  \big)^2\big[1 - f^{0}(\epsilon_{\bm{k}+\bm{q},\downarrow})\big]\\
    &\times b^{0}(\omega_{\bm{q},\alpha}) \delta\big[\epsilon_{\bm{k},\uparrow} + \omega_{\bm{q},\alpha} - \epsilon_{\bm{k} + \bm{q},\downarrow}\big] .
\end{aligned}
\end{align}
We next proceed to perform the angular part of the integral over $\bm{q}$. For a given $\bm{k}$, we then introduce a new coordinate system for $\bm{q}$ where $\theta'$ is the angle between $\bm{q}$ and the $x$-axis of the new coordinate system which is taken to be aligned with $\bm{k}$. The angle between $\bm{k}$ and the $x$-axis of the original coordinate system is denoted by $\theta$. We can then express $B^{(2)}_{\alpha}$ on the following form

\begin{align}
\begin{aligned}
    &B^{(2)}_{\alpha} = -\beta\frac{V^2 a^2}{\hbar(2\pi)^3}\!\int\!\textrm{d}\bm{k}\,f^{0}(\epsilon_{\bm{k},\uparrow})g^{e}_{\uparrow,o}(\bm{k})\!\int \!\textrm{d}q\,q\, v^{m}_{q}\nu(q)\\
    &\times \big(\Omega_A u_{q} + \Omega_B v_{q}  \big)^2  \big[1 - f^{0}(\epsilon_{k,\uparrow} + \omega_{q,\alpha})\big]b^{0}(\omega_{q,\alpha})\\
    &\times  \!\int^{2\pi}_{0}\!\textrm{d}\theta'\, \big[\cos(\theta')\cos\!\big(\theta\big) - \sin(\theta')\sin\!\big(\theta\big)\big]\\
    &\times \delta\big[\omega_{q,\alpha} - 2 h_e - 2 t (qa)(ka)\cos(\theta') - t (qa)^2\big],
\end{aligned}
\end{align}
where the square bracket of cosine factors arises because we need the component of the magnon velocity in the $x$-direction of the original coordinate system. Performing the angular integral, we obtain 

\begin{align}
\begin{aligned}
    &B^{(2)}_{\alpha} = -\beta\frac{V^2}{2t\hbar(2\pi)^3}\!\int\!\textrm{d}\bm{k}\,\frac{\cos(\theta)}{k}\,f^{0}(\epsilon_{\bm{k},\uparrow})g^{e}_{\uparrow,o}(\bm{k})\\
    &\times \!\int \!\textrm{d}q\, v^{m}_{q}\big(\Omega_A u_{q} + \Omega_B v_{q}  \big)^2  \big[1 - f^{0}(\epsilon_{k,\uparrow} + \omega_{q,\alpha})\big]\\
    &\times b^{0}(\omega_{q,\alpha})\frac{2 \,\nu(q)\Lambda_{q,\alpha,2}(k)}{\sqrt{1 - \Lambda^2_{q,\alpha,2}(k)}} \Theta(1 - |\Lambda_{q,\alpha,2}(k)|).
\end{aligned}
\end{align}
Using the assumption that the important magnon energies are significantly smaller than $k_B T$, we next approximate $f^{0}(\epsilon_{k,\uparrow} + \omega_{q,\alpha}) \approx f^{0}(\epsilon_{k,\uparrow})$. We also use that $v^{e}_k = 2 t k a^2/\hbar$, as well as $\beta [1 - f^{0}(\epsilon)]f^{0}(\epsilon) = - \partial f^{0}(\epsilon)/\partial \epsilon$. Finally, approximating loose factors of $k$ by $k_F$, motivated by the combination of Fermi-distributions, we end up with 

\begin{align}
\begin{aligned}
    &B^{(2)}_{\alpha} = \frac{-V^2}{8\pi t^2 (k_F a)^2}\!\int \!\textrm{d}q\,\big(\Omega_A u_{q} + \Omega_B v_{q}  \big)^2\\
    &\times  b^{0}(\omega_{q,\alpha}) \,v^{m}_{q} \frac{2 \,\nu(q)\Lambda_{q,\alpha,2}}{\sqrt{1 - \Lambda^2_{q,\alpha,2}}} \Theta(1 - |\Lambda_{q,\alpha,2}|)\\
    &\times\frac{1}{(2\pi)^2}\!\int\!\textrm{d}\bm{k}\,v^{e}_{k_x}\,\Bigg[-\frac{\partial f^{0}(\epsilon_{\bm{k},\uparrow})}{\partial \epsilon_{\bm{k},\uparrow}}\Bigg]g^{e}_{\uparrow,o}(\bm{k}),
\end{aligned}
\end{align}

where now 

\begin{align}
    \Lambda_{q,\alpha,2} = \frac{1}{2t(qa)(k_F a)}[\omega_{q,\alpha} - 2 h_e - t (qa)^2].
\end{align}
Writing the integral over $\bm{k}$ as a combination of electron currents and inserting $v^{m}_{q} = \kappa^2 q a^2 /(\omega_q \hbar )$, we then obtain 

\begin{align}
\begin{aligned}
    &B^{(2)}_{\alpha} = -\Gamma^{(2)}_{\alpha} \big[j_e + j_s\big], 
\end{aligned}
\end{align}
where we have defined

\begin{align}
\begin{aligned}
    &\Gamma^{(2)}_{\alpha} = \frac{1}{\hbar} \frac{(k_F a)^2 V^2}{8 \pi E^2_F} \int \!\textrm{d}(qa)\, \big(\Omega_A u_{q} + \Omega_B v_{q}  \big)^2\\
    &\times b^{0}(\omega_{q,\alpha}) \,\frac{\kappa^2 (qa)}{\omega_q} \frac{\nu(q)\Lambda_{q,\alpha,2}\,\Theta(1 - |\Lambda_{q,\alpha,2}|)}{\sqrt{1 - \Lambda^2_{q,\alpha,2}}} .
\end{aligned}
\end{align}
The integral included in $\Gamma^{(2)}_{\alpha}$ is somewhat complicated, but can, for a given set of parameters, be calculated numerically.\\ 
\indent We next evaluate $B^{(3)}_{\alpha}$, which is on the form 

\begin{align}
\begin{aligned}
    &B^{(3)}_{\alpha} = -\beta\frac{V^2 a^2}{\hbar(2\pi)^3}\! \int \!\textrm{d}\bm{q}\, v^{m}_{q_x} b^{0}(\omega_{\bm{q},\alpha})g^{m}_{\alpha,o}(\bm{q})\\
    &\times \nu(q)\big(\Omega_A u_{\bm{q}} + \Omega_B v_{\bm{q}}  \big)^2\!\int\!\textrm{d}\bm{k}\,  \big[1 - f^{0}(\epsilon_{\bm{k}+\bm{q},\downarrow})\big]\\
    &\times f^{0}(\epsilon_{\bm{k},\uparrow}) \,\delta\big[\epsilon_{\bm{k},\uparrow} + \omega_{\bm{q},\alpha} - \epsilon_{\bm{k} + \bm{q},\downarrow}\big].
\end{aligned}
\end{align}
Following similar steps as above, we proceed to, this time, perform the angular part of the integral over $\bm{k}$

\begin{align}
\begin{aligned}
    &B^{(3)}_{\alpha} = -\beta\frac{V^2 a^2}{\hbar(2\pi)^3}\! \int \!\textrm{d}\bm{q}\, v^{m}_{q_x} g^{m}_{\alpha,o}(\bm{q}) \\
    &\times \nu(q)\big(\Omega_A u_{q} + \Omega_B v_{q}  \big)^2 b^{0}(\omega_{\bm{q},\alpha})\!\int\!\textrm{d}k\,k\\
    &\times f^{0}(\epsilon_{k,\uparrow})\big[1 - f^{0}(\epsilon_{k,\uparrow} + \omega_{q,\alpha})\big]\!\int^{2\pi}_{0}\!\textrm{d}\theta'\,  \\
    &\times \delta\big[\omega_{q,\alpha} - 2 h_e - 2t (qa) (ka) \cos(\theta') - t(qa)^2 \big],
\end{aligned}
\end{align}
where we have introduced a rotated coordinate system for $\bm{k}$ where the new $x$-axis is aligned with $\bm{q}$, and $\theta'$ is the angle between $\bm{k}$ and $\bm{q}$. The result after the angular integration, becomes 

\begin{align}
\begin{aligned}
    &B^{(3)}_{\alpha} = -\beta\frac{V^2}{2t\hbar(2\pi)^3}\! \int \!\textrm{d}\bm{q}\, \frac{v^{m}_{q_x}}{q} b^{0}(\omega_{\bm{q},\alpha})g^{m}_{\alpha,o}(\bm{q})\\
    &\times \big(\Omega_A u_{q} + \Omega_B v_{q}  \big)^2\frac{2\,\nu(q)\Theta(1 - |\Lambda_{q,\alpha,3}|)}{\sqrt{1 - \Lambda^2_{q,\alpha,3}}}\\
    &\times \!\int\!\textrm{d}k\,   \big[1 - f^{0}(\epsilon_{k,\uparrow} + \omega_{q,\alpha})\big] f^{0}(\epsilon_{k,\uparrow}),
\end{aligned}
\end{align}
with 

\begin{align}
    \Lambda_{q,\alpha,3} = \frac{1}{2t(qa)(k_F a)}[\omega_{q,\alpha} - 2 h_e - t (qa)^2],
\end{align}
where we have taken $k \approx k_F$ in $\Lambda_{q,\alpha,3}$, once again motivated by the combination of Fermi distributions. We then transform the radial integral over $k$ into an integral over electron energy, producing a factor $1/\sqrt{\epsilon_{k} + \mu}$ which we approximate by its value at the Fermi level. Further, using the assumption $E_F \gg k_B T, h_e$, we end up with

\begin{align}
\begin{aligned}
    &B^{(3)}_{\alpha} = -\beta\frac{V^2}{2t\hbar(2\pi)^3}\! \int \!\textrm{d}\bm{q}\, \frac{v^{m}_{q_x}}{q} b^{0}(\omega_{\bm{q},\alpha})g^{m}_{\alpha,o}(\bm{q})\\
    &\times \big(\Omega_A u_{q} + \Omega_B v_{q}  \big)^2\frac{2\,\nu(q) \Theta(1 - |\Lambda_{q,\alpha,3}|)}{\sqrt{1 - \Lambda^2_{q,\alpha,3}}}\\
    &\times \frac{1}{2a\sqrt{t}}\frac{1}{\sqrt{E_F}}\frac{e^{\beta \omega_{q,\alpha}}}{e^{\beta \omega_{q,\alpha}} - 1}\omega_{q,\alpha}.
\end{aligned}
\end{align}
This expression can further be put on the form

\begin{align}
    &B^{(3)}_{\alpha} = \frac{-V^2}{4a t^{\frac{3}{2}}\hbar(2\pi)^3 \sqrt{E_F}}\! \int \!\textrm{d}\bm{q}\, v^{m}_{q_x}\Bigg[-\frac{\partial b^{0}(\omega_{\bm{q},\alpha})}{\partial \omega_{\bm{q},\alpha}}\Bigg] g^{m}_{\alpha,o}(\bm{q})\nonumber\\
    &\times \big(\Omega_A u_{q} + \Omega_B v_{q}  \big)^2\, \frac{\omega_{q,\alpha}}{q} \frac{2\,\nu(q)\Theta(1 - |\Lambda_{q,\alpha,3}|)}{\sqrt{1 - \Lambda^2_{q,\alpha,3}}}.
\end{align} 
The integral can here be rewritten as a combination of magnon currents multiplied by the expectation value of the second line of additional momentum dependent factors calculated using the first part of the integral as the distribution function. When the first part of the integral is a sufficiently peaked function with respect to momentum, and the second part varies slowly with momentum, this procedure approaches simply approximating the second part by its value at the momentum corresponding to the peak of the first part. Assuming that radial momentum dependence of $g^{m}_{\alpha,o}(\bm{q})$ only has a weak effect on the important momentum region for the integral, the result for $B^{(3)}_{\alpha}$ can be expressed as

\begin{align}
    &B^{(3)}_{\alpha} = - \Gamma^{(3)}_{\alpha}\big[j_m - j_{sm}\big],
\end{align}
where 

\begin{align}
    &\Gamma^{(3)}_{\alpha} = \frac{1}{\hbar}\frac{(k_F a)^3 V^2}{8\pi E^2_F} \frac{1}{I_{\alpha}} \int \!\textrm{d}(qa) (qa)\, \frac{\kappa^2 (qa)}{\omega_{q}}\Bigg[-\frac{\partial b^{0}(\omega_{\bm{q},\alpha})}{\partial \omega_{\bm{q},\alpha}}\Bigg] \nonumber\\
    &\times \big(\Omega_A u_{q} + \Omega_B v_{q}  \big)^2\, \frac{\omega_{q,\alpha}}{q a} \frac{\nu(q)\Theta(1 - |\Lambda_{q,\alpha,3}|)}{\sqrt{1 - \Lambda^2_{q,\alpha,3}}},
\end{align}
and 

\begin{align}
    I_{\alpha} = \int \!\textrm{d}(qa)\,(qa) \frac{\kappa^2 (qa)}{\omega_q}\Bigg[-\frac{\partial b^{0}(\omega_{\bm{q},\alpha})}{\partial \omega_{\bm{q},\alpha}}\Bigg].
\end{align}
Like $\Gamma^{(2)}_{\alpha}$, the expression for $\Gamma^{(3)}_{\alpha}$ can also be evaluated numerically. 

\section{Final expressions for interaction terms}\label{App:coeffs}
Performing the necessary calculations for all interaction terms, we end up with 

\begin{align}
\begin{aligned}
    &B^{(1)}_{\alpha} = \,\,\,\,\Gamma^{(1)}_{\alpha}\big[j_e - j_s\big],\\
    &B^{(2)}_{\alpha} = -\Gamma^{(2)}_{\alpha} \big[j_e + j_s\big],\\
    &B^{(3)}_{\alpha} = - \Gamma^{(3)}_{\alpha}\big[j_m - j_{sm}\big],
\end{aligned}
\end{align}

\begin{align}
\begin{aligned}
    &B^{(1)}_{\beta} = \,\,\,\,\Gamma^{(1)}_{\beta}\big[j_e + j_s\big],\\
    &B^{(2)}_{\beta} = -\Gamma^{(2)}_{\beta} \big[j_e - j_s\big],\\
    &B^{(3)}_{\beta} = - \Gamma^{(3)}_{\beta}\big[j_m + j_{sm}\big],
\end{aligned}
\end{align}

\begin{align}
\begin{aligned}
    &F^{(1)}_{\uparrow,\alpha} = \,\,\,\,\Gamma^{(1)}_{\uparrow,\alpha}\big[j_e - j_s\big],\\
    &F^{(2)}_{\uparrow,\alpha} = -\Gamma^{(2)}_{\uparrow,\alpha} \big[j_e + j_s\big],\\
    &F^{(3)}_{\uparrow,\alpha} = - \Gamma^{(3)}_{\uparrow,\alpha}\big[j_m - j_{sm}\big],
\end{aligned}
\end{align}

\begin{align}
\begin{aligned}
    &F^{(1)}_{\uparrow,\beta} = -\Gamma^{(1)}_{\uparrow,\beta}\big[j_e + j_s\big],\\
    &F^{(2)}_{\uparrow,\beta} = \,\,\,\,\Gamma^{(2)}_{\uparrow,\beta} \big[j_e - j_s\big],\\
    &F^{(3)}_{\uparrow,\beta} = \,\,\,\,\Gamma^{(3)}_{\uparrow,\beta}\big[j_m + j_{sm}\big],
\end{aligned}
\end{align}

\begin{align}
\begin{aligned}
    &F^{(1)}_{\downarrow,\alpha} = \,\,\,\,\Gamma^{(1)}_{\downarrow,\alpha}\big[j_e + j_s\big],\\
    &F^{(2)}_{\downarrow,\alpha} = -\Gamma^{(2)}_{\downarrow,\alpha} \big[j_e - j_s\big],\\
    &F^{(3)}_{\downarrow,\alpha} = \,\,\,\,\Gamma^{(3)}_{\downarrow,\alpha}\big[j_m - j_{sm}\big],
\end{aligned}
\end{align}

\begin{align}
\begin{aligned}
    &F^{(1)}_{\downarrow,\beta} = -\Gamma^{(1)}_{\downarrow,\beta}\big[j_e - j_s\big],\\
    &F^{(2)}_{\downarrow,\beta} = \,\,\,\,\Gamma^{(2)}_{\downarrow,\beta} \big[j_e + j_s\big],\\
    &F^{(3)}_{\downarrow,\beta} = -\Gamma^{(3)}_{\downarrow,\beta}\big[j_m + j_{sm}\big],
\end{aligned}
\end{align}
where 

\begin{align}
\begin{aligned}
    &\Gamma^{(1/2)}_{\alpha} = \frac{1}{\hbar} \frac{(k_F a)^2 V^2}{8 \pi E^2_F} \int \!\textrm{d}(qa)\, \big(\Omega_A u_{q} + \Omega_B v_{q}  \big)^2\\
    &\times b^{0}(\omega_{q,\alpha}) \,\frac{\kappa^2 (qa)}{\omega_q} \frac{\nu(q)\Lambda_{q,\alpha,1/2} \,\Theta(1 - |\Lambda_{q,\alpha,1/2}|)}{\sqrt{1 - \Lambda^2_{q,\alpha,1/2}}} ,
\end{aligned}
\end{align}

\begin{align}
    &\Gamma^{(3)}_{\alpha} = \frac{1}{\hbar}\frac{(k_F a)^3 V^2}{8\pi E^2_F} \frac{1}{I_{\alpha}} \int \!\textrm{d}(qa) (qa)\, \frac{\kappa^2 (qa)}{\omega_{q}}\Bigg[-\frac{\partial b^{0}(\omega_{\bm{q},\alpha})}{\partial \omega_{\bm{q},\alpha}}\Bigg] \nonumber\\
    &\times \big(\Omega_A u_{q} + \Omega_B v_{q}  \big)^2\, \frac{\omega_{q,\alpha}}{q a} \frac{\nu(q)\Theta(1 - |\Lambda_{q,\alpha,3}|)}{\sqrt{1 - \Lambda^2_{q,\alpha,3}}},
\end{align}

\begin{align}
\begin{aligned}
    &\Gamma^{(1/2)}_{\beta} = \frac{1}{\hbar} \frac{(k_F a)^2 V^2}{8 \pi E^2_F} \int \!\textrm{d}(qa)\, \big(\Omega_A v_{q} + \Omega_B u_{q}  \big)^2\\
    &\times b^{0}(\omega_{q,\beta}) \,\frac{\kappa^2 (qa)}{\omega_q} \frac{\nu(q)\Lambda_{q,\beta,1/2}\,\Theta(1 - |\Lambda_{q,\beta,1/2}|)}{\sqrt{1 - \Lambda^2_{q,\beta,1/2}}},
\end{aligned}
\end{align}

\begin{align}
    &\Gamma^{(3)}_{\beta} = \frac{1}{\hbar}\frac{(k_F a)^3 V^2}{8\pi E^2_F} \frac{1}{I_{\beta}} \int \!\textrm{d}(qa) (qa)\, \frac{\kappa^2 (qa)}{\omega_{q}}\Bigg[-\frac{\partial b^{0}(\omega_{\bm{q},\beta})}{\partial \omega_{\bm{q},\beta}}\Bigg] \nonumber\\
    &\times  \big(\Omega_A v_{q} + \Omega_B u_{q}  \big)^2\, \frac{\omega_{q,\beta}}{q a} \frac{\nu(q)\Theta(1 - |\Lambda_{q,\beta,3}|)}{\sqrt{1 - \Lambda^2_{q,\beta,3}}},
\end{align}

\begin{align}
    &\Gamma^{(1)}_{\sigma,\alpha} = \frac{1}{\hbar} \frac{(k_F a) V^2}{4 \pi E_F} \int \!\textrm{d}(qa)\, \big(\Omega_A u_{q} + \Omega_B v_{q}  \big)^2b^{0}(\omega_{q,\alpha})\nonumber \\
    &\times  \, \frac{\Theta(1 - |\Lambda_{q,\sigma,\alpha,1}|)}{\sqrt{1 - \Lambda^2_{q,\sigma,\alpha,1}}}\Bigg[1 - \frac{qa}{k_F a}\, \Lambda_{q,\sigma,\alpha,1}\Bigg],
\end{align}

\begin{align}
    &\Gamma^{(2)}_{\sigma,\alpha} = \frac{1}{\hbar} \frac{(k_F a) V^2}{4 \pi E_F} \int \!\textrm{d}(qa)\, \big(\Omega_A u_{q} + \Omega_B v_{q}  \big)^2b^{0}(\omega_{q,\alpha})\nonumber \\
    &\times  \, \frac{\Theta(1 - |\Lambda_{q,\sigma,\alpha,2}|)}{\sqrt{1 - \Lambda^2_{q,\sigma,\alpha,2}}},
\end{align}

\begin{align}
    &\Gamma^{(3)}_{\sigma,\alpha} = \frac{1}{\hbar} \frac{(k_F a)^2 V^2}{4 \pi E_F} \frac{1}{I_{\alpha}} \int \!\textrm{d}(qa) (qa) \Bigg[-\frac{\partial b^{0}(\omega_{\bm{q},\alpha})}{\partial \omega_{\bm{q},\alpha}}\Bigg] \Lambda_{q,\sigma,\alpha,3}\nonumber\\
    &\times  \big(\Omega_A u_{q} + \Omega_B v_{q}  \big)^2\, \frac{\omega_{q,\alpha}}{qa} \frac{\Theta(1 - |\Lambda_{q,\sigma,\alpha,3}|)}{\sqrt{1 - \Lambda^2_{q,\sigma,\alpha,3}}},
\end{align}

\begin{align}
    &\Gamma^{(1)}_{\sigma,\beta} = \frac{1}{\hbar} \frac{(k_F a) V^2}{4 \pi E_F} \int \!\textrm{d}(qa)\, \big(\Omega_A v_{q} + \Omega_B u_{q}  \big)^2b^{0}(\omega_{q,\beta})\nonumber \\
    &\times  \, \frac{\Theta(1 - |\Lambda_{q,\sigma,\beta,1}|)}{\sqrt{1 - \Lambda^2_{q,\sigma,\beta,1}}},
\end{align}

\begin{align}
    &\Gamma^{(2)}_{\sigma,\beta} = \frac{1}{\hbar} \frac{(k_F a) V^2}{4 \pi E_F} \int \!\textrm{d}(qa)\, \big(\Omega_A v_{q} + \Omega_B u_{q}  \big)^2 b^{0}(\omega_{q,\beta})\nonumber \\
    &\times  \, \frac{\Theta(1 - |\Lambda_{q,\sigma,\beta,2}|)}{\sqrt{1 - \Lambda^2_{q,\sigma,\beta,2}}}\Bigg[1 - \frac{qa}{k_F a}\, \Lambda_{q,\sigma,\beta,2}\Bigg],
\end{align}

\begin{align}
    &\Gamma^{(3)}_{\sigma,\beta} = \frac{1}{\hbar} \frac{(k_F a)^2 V^2}{4 \pi E_F} \frac{1}{I_{\beta}} \int \!\textrm{d}(qa) (qa) \Bigg[-\frac{\partial b^{0}(\omega_{\bm{q},\beta})}{\partial \omega_{\bm{q},\beta}}\Bigg] \Lambda_{q,\sigma,\beta,3}\nonumber\\
    &\times  \big(\Omega_A v_{q} + \Omega_B u_{q}  \big)^2\, \frac{\omega_{q,\beta}}{qa} \frac{\Theta(1 - |\Lambda_{q,\sigma,\beta,3}|)}{\sqrt{1 - \Lambda^2_{q,\sigma,\beta,3}}}.
\end{align}
\noindent We have here defined 

\begin{align}
    I_{\gamma} = \int \!\textrm{d}(qa)\,(qa)\, \frac{\kappa^2 (qa)}{\omega_q}\Bigg[-\frac{\partial b^{0}(\omega_{\bm{q},\gamma})}{\partial \omega_{\bm{q},\gamma}}\Bigg],
\end{align}
as well as

\begin{align}
\begin{aligned}
    &\Lambda_{q,\gamma,1} = \frac{1}{2t(qa)(k_F a)}[\omega_{q,\gamma} + 2 \gamma h_e + t (qa)^2],\\
    &\Lambda_{q,\gamma,2} = \frac{1}{2t(qa)(k_F a)}[\omega_{q,\gamma} + 2 \gamma h_e - t (qa)^2],\\
    &\Lambda_{q,\gamma,3} = \Lambda_{q,\gamma,2},
\end{aligned}
\end{align}

\begin{align}
\begin{aligned}
    &\Lambda_{q,\uparrow,\alpha,1} = \frac{1}{2t(qa)(k_F a)}[\omega_{q,\alpha} - 2 h_e + t (qa)^2],\\
    &\Lambda_{q,\uparrow,\alpha,2} = \frac{1}{2t(qa)(k_F a)}[\omega_{q,\alpha} - 2 h_e - t (qa)^2],\\
    &\Lambda_{q,\uparrow,\alpha,3} = \Lambda_{q,\uparrow,\alpha,2},\\
    &\Lambda_{q,\uparrow,\beta,1} = \frac{1}{2t(qa)(k_F a)}[\omega_{q,\beta} + 2 h_e + t (qa)^2],\\
    &\Lambda_{q,\uparrow,\beta,2} = \frac{-1}{2t(qa)(k_F a)}[\omega_{q,\beta} + 2 h_e - t (qa)^2],\\
    &\Lambda_{q,\uparrow,\beta,3} = \Lambda_{q,\uparrow,\beta,1},
\end{aligned}
\end{align}

\begin{align}
\begin{aligned}
    &\Lambda_{q,\downarrow,\alpha,1} = - \Lambda_{q,\uparrow,\alpha,2},\\
    &\Lambda_{q,\downarrow,\alpha,2} = - \Lambda_{q,\uparrow,\alpha,1},\\
    &\Lambda_{q,\downarrow,\alpha,3} = \Lambda_{q,\uparrow,\alpha,1},\\
    &\Lambda_{q,\downarrow,\beta,1} = - \Lambda_{q,\uparrow,\beta,2},\\
    &\Lambda_{q,\downarrow,\beta,2} = \Lambda_{q,\uparrow,\beta,1},\\
    &\Lambda_{q,\downarrow,\beta,3} = - \Lambda_{q,\uparrow,\beta,2}.
\end{aligned}
\end{align}

\section{Definition of coefficients}\label{App:expressions}
The coefficients in Eq.\! \eqref{eq:rightarrow_s} are given by

\begin{align}
\begin{aligned}
    &A_{e\rightarrow s} = \frac{1}{X_s}\Bigg[\frac{1}{2} \Big(Y_0 \frac{T_{-}}{T_{+}} + P_0\Big)\tau^{-1}_e\\
    &+ \Big(1 + \frac{T_{-}}{T_{+}}\Big)\Big(- \Gamma^{(1)}_{\downarrow,\alpha} + \Gamma^{(2)}_{\downarrow,\alpha} + \Gamma^{(1)}_{\downarrow,\beta} - \Gamma^{(2)}_{\downarrow,\beta} \Big)\\
    &+ \Big(1 - \frac{T_{-}}{T_{+}}\Big)\Big(\Gamma^{(1)}_{\uparrow,\alpha} - \Gamma^{(2)}_{\uparrow,\alpha} - \Gamma^{(1)}_{\uparrow,\beta} + \Gamma^{(2)}_{\uparrow,\beta} \Big)\Bigg],
\end{aligned}
\end{align}

\begin{align}
\begin{aligned}
    &A_{m \rightarrow s} = \frac{1}{X_s}\Bigg[\Big(1 - \frac{T_{-}}{T_{+}}\Big)\Big(\Gamma^{(3)}_{\uparrow,\beta} - \Gamma^{(3)}_{\uparrow,\alpha}  \Big)\\
    &+ \Big(1 + \frac{T_{-}}{T_{+}}\Big)\Big(- \Gamma^{(3)}_{\downarrow,\alpha} + \Gamma^{(3)}_{\downarrow,\beta}  \Big)\Bigg],
\end{aligned}
\end{align}
and

\begin{align}
\begin{aligned}
    &A_{sm \rightarrow s} = \frac{1}{X_s}\Bigg[\Big(1 - \frac{T_{-}}{T_{+}}\Big)\Big(\Gamma^{(3)}_{\uparrow,\beta} + \Gamma^{(3)}_{\uparrow,\alpha}  \Big)\\ 
     &+ \Big(1 + \frac{T_{-}}{T_{+}}\Big)\Big(\Gamma^{(3)}_{\downarrow,\alpha} + \Gamma^{(3)}_{\downarrow,\beta}  \Big) \Bigg],
\end{aligned}
\end{align}
where

\begin{align}
\begin{aligned}
    &X_s = \frac{1}{2}\Big(P_0 \frac{T_{-}}{T_{+}} + Y_0\Big) \tau^{-1}_e\\
    &-\Big(\frac{T_{-}}{T_{+}} - 1 \Big)\Big(\Gamma^{(1)}_{\uparrow,\alpha} + \Gamma^{(2)}_{\uparrow,\alpha} +  \Gamma^{(1)}_{\uparrow,\beta} + \Gamma^{(2)}_{\uparrow,\beta} \Big)\\
    &+ \Big(\frac{T_{-}}{T_{+}} + 1 \Big)\Big(\Gamma^{(1)}_{\downarrow,\alpha} + \Gamma^{(2)}_{\downarrow,\alpha} + \Gamma^{(1)}_{\downarrow,\beta} + \Gamma^{(2)}_{\downarrow,\beta} \Big). 
\end{aligned}
\end{align}
Further, the coefficients in Eq.\! \eqref{eq:rightarrow_m} take the form

\begin{align}
\begin{aligned}
    C_{e \rightarrow m} &= A_{e\rightarrow m} + A_{e\rightarrow s}\,A_{s\rightarrow m},\\
    C_{sm \rightarrow m} &= A_{sm\rightarrow m} + A_{sm\rightarrow s}\, A_{s\rightarrow m},
\end{aligned}
\end{align}
where 

\begin{align}
\begin{aligned}
    &A_{e\rightarrow m} = \big[\Gamma^{(1)}_{\beta} + \Gamma^{(1)}_{\alpha} - \Gamma^{(2)}_{\beta} - \Gamma^{(2)}_{\alpha}\big]/X_m,\\
    &A_{s\rightarrow m} = \big[\Gamma^{(1)}_{\beta} - \Gamma^{(1)}_{\alpha} + \Gamma^{(2)}_{\beta} - \Gamma^{(2)}_{\alpha}\big]/X_m,\\
    &A_{sm\rightarrow m} = \big[\Gamma^{(3)}_{\alpha} - \Gamma^{(3)}_{\beta}\big]/X_m,
\end{aligned}
\end{align}
and 

\begin{align}
\begin{aligned}
    &X_{m} = \tau^{-1}_{M_0} + \big[\Gamma^{(3)}_{\beta} + \Gamma^{(3)}_{\alpha} \big]\\
    &- A_{m\rightarrow s}\big[\Gamma^{(1)}_{\beta} - \Gamma^{(1)}_{\alpha} + \Gamma^{(2)}_{\beta} - \Gamma^{(2)}_{\alpha}\big].
\end{aligned}
\end{align}


\bibliography{Refs}
\end{document}